\documentclass[twocolumn]{aastex62}

\usepackage{graphicx}
\usepackage{amssymb}
\usepackage{xspace}

\newcommand{\Sersic}{S\'{e}rsic }

\newcommand{\dev}{de Vaucouleurs' }

\begin{document}
\graphicspath{{./}{figures/}}

\title{Evolution Through the Post-Starburst Phase: \\ Using Post-Starburst Galaxies as Laboratories for Understanding the Processes that Drive Galaxy Evolution}
\author[0000-0002-4235-7337]{K. Decker French}
\affil{Department of Astronomy, University of Illinois, 1002 W. Green St., Urbana, IL 61801, USA}

\begin{abstract}
Post-starburst (or ``E+A") galaxies trace the fastest and most dramatic processes in galaxy evolution. Recent work studying the evolution of galaxies through this phase have revealed insights on how galaxies undergo structural and stellar population changes as well as the role of various feedback mechanisms. In this review, I summarize recent work on identifying post-starburst galaxies; tracing the role of this phase through cosmic time; measuring stellar populations, on-going star formation, morphologies, kinematics, interstellar medium properties, and AGN activity; mechanisms to cause the recent starburst and its end; and the future evolution to quiescence (or not). The review concludes with a list of open questions and exciting possibilities for future facilities. 
\end{abstract}

\section{Introduction}

Observational surveys of galaxies reveal two broad classes: star-forming and quiescent. This bimodality in color extends to star-formation properties, gas properties, kinematics, morphologies, and connects to mean shifts in stellar mass and absolute magnitude. As our understanding of this bimodality has grown to include the idea that star-forming galaxies eventually stop forming new stars at a significant rate and become quiescent, the question of how galaxies cease their star formation has been raised.

Post-starburst (or ``E+A'', ``K+A"\footnote{While the terms are sometimes used interchangeably, they refer to the observed spectral signatures of these galaxies which display a combination of young (A-star) and old (elliptical or K-star) populations. The details of various definitions are discussed further in \S\ref{identifying}.}) galaxies are caught in a rapid transition between these classes of star-forming and quiescent. The question of how best to observationally identify post-starburst galaxies is discussed in detail in \S\ref{identifying}, but these galaxies are often identified via a lack of nebular emission lines (signifying a low current star formation rate) together with strong Balmer absorption. Strong Balmer absorption is indicative of a substantial population of A stars, indicating these galaxies have experienced a burst of star formation sometime in the past billion years.

Spectroscopic observations of local galaxies show several examples of galaxies with unusual, A-star dominated spectra, especially in surveys of interacting galaxies\footnote{\citet{Arp1969} identified the A-star dominated spectra of several interacting galaxies, including NGC 5195 (M51b), although incorrectly attributed this scenario to expulsion rather than the merging of these galaxies.}. Alongside advancements in the understanding of galaxy-galaxy mergers and their connection to the observed range in tidal features and disturbed morphologies \citep{Toomre1972} and in stellar population synthesis \citep{Tinsley1968,Larson1978,Tinsley1979}, the interpretation of A-star dominated spectra as originating from young, composite, post-starburst, stellar populations emerged \citep[e.g.,][]{Spinrad1973,Schweizer1978, Schweizer1982, Dressler1983,Couch1987}. Large spectroscopic surveys such as the LCRS \citep{Shectman1996}
and SDSS \citep{Strauss2002} have allowed for the study of post-starburst galaxies as a population \citep[e.g.,][]{Zabludoff1996, Goto2005}.

Modern facilities have allowed for more detailed studies of post-starburst galaxies from the local universe to high redshift, solving some mysteries and deepening others. In this review,
I discuss the current state of understanding on selecting post-starburst galaxies in \S2, 
the post-starburst phase through cosmic time in \S3, 
characterizing their stellar populations in \S4, 
constraining current star formation rates in \S5, 
morphologies and stellar population gradients in \S6, 
kinematics and dynamics in \S7, 
the interstellar medium in \S8, 
and AGN and nuclear activity in \S9. 
A synthesis of our understanding of the past history of these galaxies is discussed in \S10, 
and the future evolution of these galaxies in \S11. 
In \S12, I conclude and present a list of open questions in the field.

\section{Identifying post-starburst galaxies}
\label{identifying}

\begin{figure*}[t!]
\centering
\includegraphics[width=\textwidth]{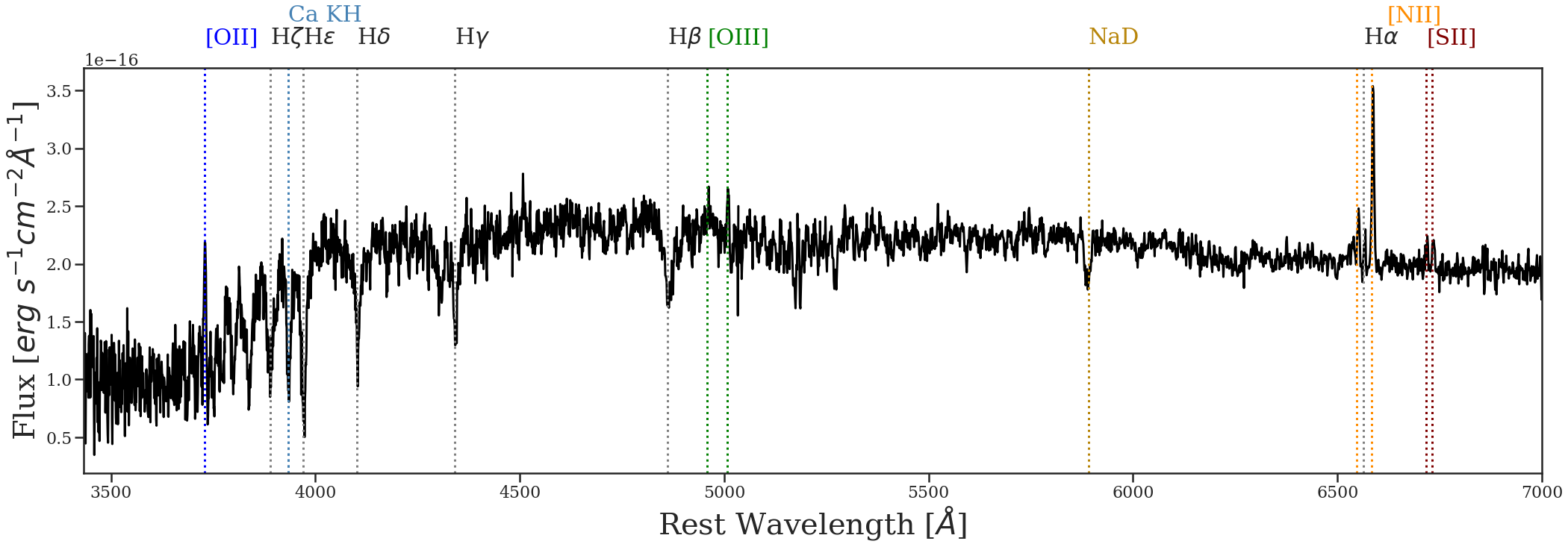}
\caption{Example spectrum of a post-starburst galaxy from the SDSS (spec-0524-52027-0492). This galaxy was selected by all three of the \citet{Wild2007}, \citet{Alatalo2016}, and \citet{French2018} selection methods. This galaxy has strong Balmer absorption and weak emission lines that are more consistent with shock or AGN-dominated ionization than ionization from star formation. In this spectrum, the H$\epsilon$ line is blended with the Ca H line, resulting in stronger emission than Ca K, a quick visual diagnostic for post-starburst spectra.}
\label{fig:example}
\end{figure*}

Some of the first systematic searches for post-starburst galaxies were in spectroscopic surveys of clusters, where a subset of galaxies was observed to have strong Balmer absorption lines like star-forming galaxies, yet with missing or weak emission lines from on-going star formation \citep{Dressler1983,Couch1987}. A typical post-starburst optical spectrum is shown in Figure \ref{fig:example} from the SDSS \citep{York2000, Strauss2002}. These early studies have motivated systematic efforts to select post-starburst galaxies from large spectroscopic surveys. A summary of the methods discussed in this section is shown in Table \ref{tab:selection}. 

In order to select for recent star formation, post-starburst selection methods typically require strong Balmer absorption, especially in H$\beta\gamma\delta\epsilon$. The H$\delta$ line is often used due to the smooth continuum region around the line and the relative lack of nebular emission filling. The Lick H$\delta_{\rm A}$ and H$\gamma_{\rm A}$ indices \citep{Worthey1997} are optimized for the continuum region and expected widths of these lines in A stars. Post-starburst selection methods typically make use of a combination of Balmer lines \citep{Zabludoff1996,Brown2009} or the H$\delta$ line \citep{Dressler1999,Goto2005,French2015,Alatalo2016}. Another approach is to fit spectral templates to galaxy spectra, and select galaxies with preferentially more light from young $\sim$ Gyr old templates than older K-star dominated templates \citep{Quintero2004,Mendel2013,Ciesla2016, Pattarakijwanich2016}. \citet{Wild2007,Wild2009} define a post-starburst selection using a principal component analysis (PCA) to the region around rest-frame 4000 \AA, finding one of the PCA components to be closely correlated with Lick H$\delta_{\rm A}$. 

To select against current star formation, many post-starburst selection methods require low levels of emission line flux. The EW [OII]$\lambda3727$ is often used due to its proximity to the Balmer lines and ease of use in surveys with limited spectral coverage \citep{Zabludoff1996,Dressler1999,Goto2005}. Surveys with wider spectral coverage like the SDSS provide access to another commonly used indicator H$\alpha$ \citep{Quintero2004,Goto2005,Brown2009,French2015}. The use of H$\alpha$ reduces the inadvertent selection against AGN (especially LINERs) \citep{Yan2006}, although a cut on H$\alpha$ emission still selects against galaxies with strong narrow line AGN as well as galaxies with strong shocks (as might be expected after a merger). Selecting on H$\alpha$ instead of [OII]$\lambda3727$ also reduces contamination from dusty star-forming galaxies. Several methods allow for stronger emission, with selection against on-going star formation using color \citep{Yesuf2014} or by allowing for emission only if a galaxy has emission line ratios more consistent with shocks than star-formation \citep{Alatalo2016}. The PCA selection of \citet{Wild2007,Wild2009} has a component closely correlated with D$_n4000$, which selects against some on-going star formation while allowing for AGN to still enter the sample. In practice, discriminating between star formation and AGN activity in post-starburst galaxies is complex, and samples such as \citet{Wild2007,Wild2009,Alatalo2016} contain galaxies with higher SFRs and younger stellar populations than the other samples considered here (see \S\ref{sec:stellarpops}. Emission line ratios can also be used to find galaxies just after their starbursts have begun to end, before they would meet many of the post-starburst selection methods considered here \citep{Citro2017}. Recent theoretical work by \citet{Zheng2020} has shown that the H$\alpha$ flux can vary during the post-starburst phase, further complicating these post-starburst selection methods.

Several methods have been developed to select post-starburst galaxies via unsupervised machine learning methods. \citet{Meusinger2016} find a cluster of galaxies similar to post-starburst galaxies using self organized maps, using an artificial neutral network on the SDSS spectroscopic sample. The \citet{Goto2007} catalog was used to identify clusters of post-starburst galaxies to select a larger sample of 2665 galaxies with similar properties. \citet{Baron2017b} use an outlier detection method, training an unsupervised random forest algorithm on the SDSS spectroscopic sample, which finds many classes of unusual objects, including post-starburst galaxies. In the sample of the 400 galaxies with the highest outlier scores, 33 have strong H$\delta$ absorption. Many of these galaxies are in the previous post-starburst samples discussed above, but several show complex emission line profiles that would have excluded them from some previous samples due to the possibility of on-going star formation.

Several photometric selection methods are also used for post-starburst galaxies. At high redshift, quiescent galaxies are often selected using the UVJ diagram, with post-starburst galaxies populating a unique position in this space \citep{Whitaker2012,Yano2016,Suess2020}. Based on the PCA selection of \citet{Wild2007,Wild2009}, a super color method \citep{Wild2014,Wild2016} has also been used to select higher redshift post-starburst galaxies with high accuracy compared to spectroscopic selection \citep{Maltby2016}. At lower redshift, UV-IR photometry from large surveys can be used to select possible post-starburst galaxies with machine learning methods trained on spectroscopic samples \citep{French2018c}.

\begin{table*}[t!]
    \centering
    \begin{tabular}{lll}
    \hline
    Reference & Selection against current SF$^a$ & Selection for recent burst$^b$  \\
    \hline
    \citet{Zabludoff1996} & EW [OII]$\lambda3727 \ <2.5$   &    $\langle H\delta\gamma\beta\rangle > 5.5$  \\
    \citet{Dressler1999} ``a+k" & [OII]$\lambda3727$ absent & EW H$\delta \ge 8$  \\
    \citet{Dressler1999} ``k+a" & [OII]$\lambda3727$ absent & $3 \ge$ EW H$\delta < 8$  \\
    \citet{Quintero2004} & EW H$\alpha$ &  A/K ratio; H$\alpha$/(A/K) $<5$; A/K $>$0.2 \\
    \citet{Goto2005,Goto2007}$^c$ &  EW [OII]$\lambda3727 \ <2.5$ ; H$\alpha < 3$  & EW H$\delta >5$  \\
    \citet{Wild2007,Wild2009} & PCA selection ($\approx D_n4000$) & PCA selection 
    ($\approx$ Lick H$\delta_{\rm A}$) \\
    \citet{Brown2009} & log(H$\alpha$ EW) $<0.2\times$ Lick (H$\delta_{\rm A}$ + H$\gamma_{\rm A}$)/2 &
    (H$\delta_{\rm A}$ + H$\gamma_{\rm A}$)/2 $>3$ \\
    \citet{French2015,French2018} & H$\alpha < 3$  & Lick H$\delta_{\rm A}$ $-$ $\sigma$(H$\delta_{\rm A}$) $>$ 4\\
    \citet{Alatalo2016} & Shocked emission line ratios & EW H$\delta$ $>5$  \\
    \citet{Baron2017b,Baron2021} & none/outlier score/AGN-like$^d$ & EW H$\delta$ $>5$ \\
    \hline
    \end{tabular}
    \caption{Summary of post-starburst selection methods described in \S\ref{identifying} used to select large samples of post-starburst galaxies from spectroscopic surveys. We follow the convention commonly used in this literature where the positive values of EW are used to represent emission in column $a$ and absorption in column $b$. All units are in \AA\ except ratios (A/K). $^c$ \citet{Meusinger2016} use this sample to inform their selection using self-organized maps. $^d$ Multiple sample selections are included in \citet{Baron2021}, some are selected to have AGN-like line ratios, others have no cut against on-going star formation and are instead selected to be similar to a galaxy with ionized outflows are strong Balmer absorption.}
    \label{tab:selection}
\end{table*}

\section{The post-starburst phase traced through cosmic time}
\label{sec:cosmictime}

The fraction of post-starburst galaxies increases with redshift, from $<1$\% of the total galaxy population at $z\sim0$ to $>5$\% at $z\sim2$ \citep{Wild2016}. At high redshifts $z>2$, massive quiescent galaxies often show post-starburst signatures, with 34\% of quiescent galaxies being post-starburst at $z\sim2.5$ \citep{Belli2019}. At $z\sim3$, most quiescent galaxies show post-starburst signatures; a stacking analysis of 9 massive $z\sim3$ quiescent galaxies by \citet{DEugenio2020b} show the average spectrum to be post-starburst. 

Post-starburst galaxies at high redshift may be the descendants of sub-millimeter galaxies (SMGs); massive quiescent galaxies at $z\sim2$ are matched in mass and size distributions to $z>3$ SMGs \citep{Toft2014}. The rapid end of the starburst in SMGs would result in a traditional post-starburst signature \citep{Wild2020}. The large scale clustering of post-starburst galaxies from $0.5<z<3$ is consistent with SMGs \citep{Wilkinson2021}.

How many galaxies evolve to quiescence rapidly, through the post-starburst phase, vs. slowly over many Gyr? The relative importance of fast vs. slow processes appears to depend on redshift, with faster processes more common at high redshift. The growth rate of quiescent galaxies is consistent with higher fraction of post-starburst galaxies as redshift increases \citep{Whitaker2012,Wild2016,Rowlands2017,Belli2019}. These analyses are sensitive to how post-starburst galaxies are defined, what their observability timescales are, and whether growth of the quiescent galaxy population is ``one-way." \citet{Belli2019} find that at $z\sim2$, the growth rate of quiescent galaxies is twice that of post-starburst galaxies, with less growth from galaxies transitioning through the post-starburst phase as redshift decreases. \citet{Wild2020} find that at $z\sim1$, 25--50\% of the growth rate of quiescent galaxies is caused by rapid evolution through the post-starburst phase. Integrated to $z=0$ roughly half of quiescent galaxies likely experienced a period of rapid evolution through the post-starburst phase \citep{Wild2009,Snyder2011,Wild2016}, indicating the importance of this phase despite its brief duration.

The mass of post-starburst galaxies depends on redshift, with higher stellar masses found for post-starburst galaxies selected at high redshift compared to low redshift samples \citep{Wong2012,Wild2016,Almaini2017}, consistent with expectations for downsizing, where higher mass galaxies evolve to quiescence more rapidly than low mass galaxies. However, a full comparison is difficult due to the low luminosity of quiescent galaxies at stellar masses $<10^{10}$ M$_\odot$ at $z>2$, and the varying selection methods that can be used. 

The mechanisms that trigger the high redshift starbursts may be different than low redshift starbursts; high gas densities at $z>2$ can result in rapid compaction and a burst of star formation \citep{Zolotov2015}. The influence of this phase in driving rapid evolution may account for the increased fraction of post-starburst galaxies at high redshift, in combination with the decreased time available for galaxy formation when the universe was younger. The evolution of the post-starburst stellar mass function with redshift may be related to the different starburst triggering mechanisms, with higher mass galaxies ending star formation after episodes of compaction and lower mass galaxies ending star formation after mergers and interactions \citep{Wild2016}. However, these mechanisms may not be fully independent, if mergers and interactions are the initiators for episodes of compaction \citep{Zolotov2015}.

\section{Stellar populations}
\label{sec:stellarpops}

The current stellar populations of galaxies reveal the integrated past history of star formation. Modeling the stellar populations using stellar population synthesis (SPS) models is a powerful tool, yet requires many assumptions. The post-starburst selection methods described in \S\ref{identifying} typically find galaxies with starbursts $\sim10$ Myr to 1 Gyr ago (consistent with the lifetimes of A stars by selection). The UV-optical spectral energy distributions (SEDs) are dominated by the younger stellar populations. By mass, the recent starbursts contributed a few percent to tens of percent of the current stellar mass in low redshift samples \citep{Kaviraj2007a,Du2010a,Melnick2013,French2018}, and 40-90\% in a sample of $z\sim1$ post-starbursts \citep{Wild2020}. At low redshift, galaxies evolving through the post-starburst phase have experienced a recent starburst on top of an underlying old stellar population, but at higher redshifts, the recent starbursts appear to be responsible for forming a large portion of the galaxy.

The colors of post-starburst galaxies, as well as their positions in [OIII]-H$\delta$, H$\alpha$-H$\delta$, or other spaces used to select them (\S\ref{identifying}), are degenerate in the age of the starburst and the fraction of stellar mass produced in it. This is known as the age-burst degeneracy \citep{Liu1996,Leonardi1996}, where weaker but younger starbursts will have a similar effect on the appearance of a post-starburst galaxy as stronger but older starbursts. Using spectral information over a larger wavelength range, especially in the blue and UV, helps to break this degeneracy. Because of this, single spectral indices make poor indicators of post-starburst age, as they are sensitive to the light-weighted age of the average stellar population. Colors or pairs of colors are subject to the same degeneracies, in addition to uncertainties in metallicity and dust.

Because younger stellar populations have lower mass to light ratios, the characteristics of the youngest stars can be most robustly determined. A 200 Myr stellar population will look very different than a 400 Myr population; while a 5 Gyr and 10 Gyr old quiescent galaxy will look very similar. This complicates the inference of star formation histories for many galaxies \citep[see][]{Conroy2013}, but means that the period of most recent star formation in post-starburst galaxies can be well determined. Since the youngest stars will dominate the light, the age since the starburst \textit{ended} can be most easily measured in post-starburst galaxies, with the age since the starburst began and the duration of the starburst more difficult to constrain \citep{French2018}.

In order to display an E+A signature and be selected into the criteria discussed in \S\ref{identifying}, galaxies must have experienced rapid declines in the recent starburst, on timescales $<100-200$ Myr \citep{Wild2009,French2018} These short timescales are indeed observed in stellar population fitting of post-starburst galaxies \citep{Kaviraj2007a,Wild2009,French2018,Wild2020,Forrest2020}. A burst that declines on a $>200$ Myr timescale will not have a strong enough Balmer absorption signature after the galaxy is quiescent, even if a sizable fraction of stellar mass were produced. Depending on the post-starburst selection method, galaxies experiencing a recent sharp truncation of star formation instead of a burst can display spectral signatures similar to post-starburst galaxies \citep{Ciesla2016, Pawlik2019}, although the strongest H$\delta$ absorption galaxies can only arise from a bursty recent star formation history \citep{Leonardi1996, Poggianti2009}.

Reconstructing the past histories of post-starburst galaxies allows for their maximum burst-phase SFRs to be compared with populations of starbursting galaxies. Stellar population fitting in post-starburst galaxies find the recent starbursts to have maximum SFRs of $10-1000$ M$_\odot$ yr$^{-1}$ \citep{Kaviraj2007a,French2018,Wild2020,Forrest2020}, and sSFRs $10-100\times$ above the main sequence of star-forming galaxies, consistent with populations of starburst galaxies. However, these estimates are subject to uncertainties in the burst duration, which can be difficult to constrain.

The unique star formation histories of post-starburst galaxies make them a useful test of stellar population models. The flux from TP-AGB stars peaks during this phase, and comparing post-starburst SEDs to model SEDs in the near-IR can constrain models with varying influence of TP-AGB stars \citep{Kriek2010,Conroy2010, Zibetti2013}. Post-starburst SEDs provide useful test cases for stellar population fitting codes to showcase the flexibility of allowed star formation histories \citep{Johnson2017}.

The recent star formation histories of post-starburst galaxies can also be constrained using star cluster measurements. \citet{Yang2008} use \textit{HST} observations of newly-formed star clusters to constrain the starburst ages of four post-starburst galaxies with sufficiently bright star clusters. Chandar et al. (2021, in press) used \textit{HST} observations of a post-starburst galaxy with deep multi-band imaging extending into the blue to constrain the star formation history of this galaxy over multiple age bins. The star cluster inferred post-burst age, burst duration, and peak star formation rate are consistent with those inferred from the integrated light modeling, with differences attributed to dust modeling and aperture effects.

\section{Measuring Star Formation Rates}
\label{sec:sfr}

Despite the selection against on-going star formation present in most post-starburst selection methods, \textit{measuring} current star formation rates (SFRs) is complicated in practice by (1) uncertainties on the dust geometry, (2) the long duration of many tracers relative to the time post-starbursts have been quiescent, and (3) the presence of AGN activity.

Nebular and forbidden emission lines trace star formation on short time scales ($<10$ Myr) including H$\alpha$ or [OII] in the optical, the Paschen or Brackett lines in the near IR, or [Ne II] and [Ne III] in the mid IR. These lines can be contaminated with emission from AGN, LINERs, or shocks. If enough emission lines are measurable, AGN contamination can be removed, as in \citet{Wild2010}. The AGN contribution to emission lines is often easier to remove in these galaxies than the AGN contribution to the 1.4 GHz line (see above), TIR, or SED. Dust attenuation can be corrected using the expected ratios between the hydrogen emission lines, although this requires the detection of multiple lines, which can be challenging due to the intrinsically weak emission lines and the need to correct for strong Balmer absorption in post-starburst galaxies. Correcting for dust attenuation also requires the assumption of a dust geometry, and may miss the presence of a heavily obscured core \citep{Smercina2018}. Observations of infrared SFR tracers with \textit{JWST} could help to calibrate the typical amounts of contamination and scatter introduced by the above effects.

The possibility of dust obscuration as the cause of the unusual E+A spectral signature has been raised by a number of studies \citet{Smail1999, Poggianti2000}. \citet{Smail1999} observed 1.4 GHz radio emission in a sample of 5 post-starbursts in a $z\sim0.4$ cluster, implying SFRs of order $20-200$ M$_\odot$yr$^{-1}$. Radio continuum emission at 1.4 GHz can be used to trace star-formation, tracing synchrotron radiation from cosmic rays accelerated by supernova remnants, and is insensitive to dust obscuration. Further studies of the 1.4 GHz emission in post-starburst galaxies have not found such high SFRs. \citet{Miller2001} observed high radio 1.4 GHz - traced SFRs in 2/15 post-starburst galaxies considered from the \citet{Zabludoff1996} sample, although at much lower levels ($\sim 2-5$ M$_\odot$yr$^{-1}$) than \citet{Smail1999}. Limits on the SFRs for the remaining 13/15 galaxies were $\lesssim1.5$ M$_\odot$yr$^{-1}$. In a larger study of 811 galaxies using an updated version of the \citet{Goto2007} catalog, \citet{Nielsen2012} found an average upper limit of 1.4 GHz derived SFRs of $<1.6$ M$_\odot$yr$^{-1}$ in a stacking analysis. This flux is dominated by the $\sim4$\% of the sample with detectable 1.4 GHz emission, which may even be from AGN. The presence of AGN or LINERs (see \S\ref{sec:agn}) can contaminate the 1.4 GHz continuum, resulting in an overestimate of the true current SFR. \citet{Moric2010} finds the 1.4 GHz continuum to be a poor tracer of star formation in BPT-selected LINERs. \citet{Moric2010} compare the 1.4 GHz inferred SFRs to SFRs inferred from NUV-NIR SED fitting, and find no significant correlation for the LINER galaxies, with the 1.4 GHz SFRs 1-100$\times$ higher than the NUV-NIR SFRs. Given that the 1.4 GHz SFRs are likely to be an overestimate of the true SFRs in post-starbursts, the lack of high inferred SFRs for 96\% \citep{Nielsen2012} imply heavy dust obscuration is uncommon in post-starburst galaxies with selections similar to \citet{Goto2007}. Some post-starburst galaxies appear to have dense concentrations of dust at their centers, yet still have low current SFRs and evidence of steep drops in SFR from their starburst peaks \citep{Smercina2018}.

SFR tracers sensitive to star formation on timescales of $>100$ Myr are contaminated by the recent burst in post-starburst galaxies. The total infrared (TIR) luminosity is especially susceptible to the unusual radiation field present from the A star dominated stellar populations, and overestimates the true SFR. Simulations of the TIR luminosity in galaxies during and after a starburst by \citet{Hayward2014} find the TIR luminosity to overestimate the instantaneous SFR by $\sim30-100\times$ during the post-starburst phase.  \citet{Smercina2018} compare TIR based SFRs to SFRs from [Ne II] and [Ne III] emission, finding the TIR based SFRs to be higher by $\sim3-4\times$. \citet{Baron2021} measured TIR-based SFRs for a variety of post-starburst samples. Those selected using a cut against H$\alpha$ emission \citep{French2018} had negligible TIR SFRs, while those selected to allow for non-SF-like emission line ratios consisted of 13-45\% galaxies above the main sequence of star formation. Modeling of the possible contribution of A-star heating using Starburst99 \citep{sb99} was not sufficient to drive the TIR luminosities in these galaxies. However, we note that samples of young post-starburst galaxies may be especially susceptible to overestimated SFRs when longer-duration SFR tracers are used.

Spectral energy distribution (SED) fitting to synthetic stellar populations of UV-optical or UV-IR photometry are often used to measure current star formation rates, but these can be heavily contaminated by the recent starburst depending on what assumptions are made for the star formation history. Even if the star formation history is flexible enough to accommodate a recently-ended starburst, the total light will still be dominated by the recent burst (by selection), and the remaining contribution from on-going star formation can be difficult to measure.

\section{Morphologies}

\begin{figure*}
    \centering
    \includegraphics[width=\textwidth]{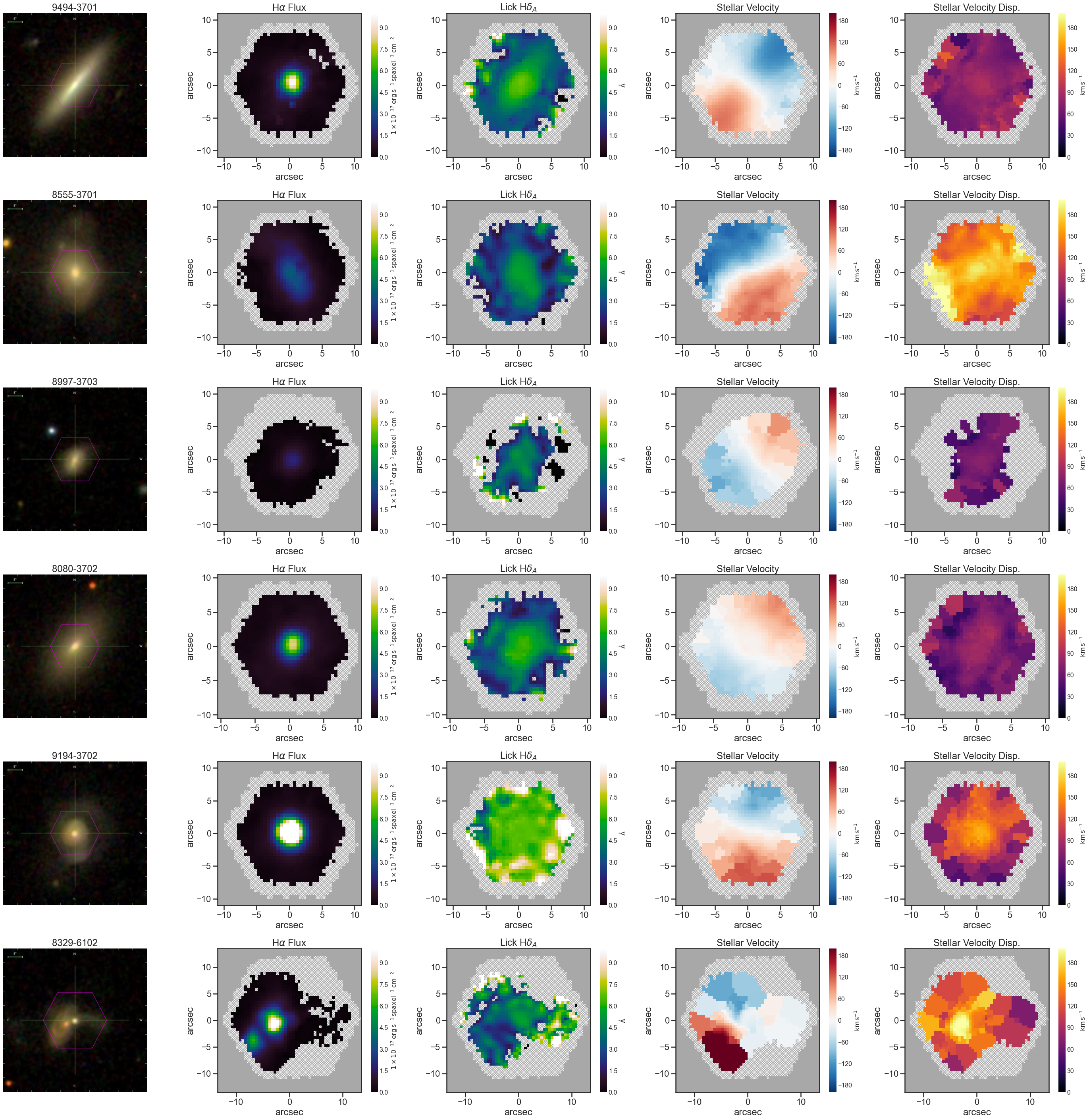}
    \caption{SDSS images and selected MaNGA observations \citep{manga,marvin} of six post-starburst galaxies from a variety of samples. The left-most column shows the SDSS $gri$ postage stamp images of each galaxy, with the MaNGA field of view overlaid in purple. Each galaxy is labeled with its MaNGA {\tt plateifu} indicator. The galaxies 9494-3701 and 8555-3701 are from the \citet{French2015,French2018} sample, the galaxy 8997-3703 is from both the \citet{Wild2010,French2018} samples, the galaxy 8080-3702 is from the \citet{Alatalo2016} sample, and the galaxies 9194-3702 and 8329-6102 are from the \citet{Wild2010} sample. The subsequent columns show the MaNGA observations of the H$\alpha$ flux, the Lick H$\delta_{\rm A}$ index, the stellar velocity field, and the stellar velocity dispersion. Each colorbar has the same range for each galaxy. 
    }
    \label{fig:manga}
\end{figure*}

\subsection{Tidal Features and Disturbed Morphologies in Low Redshift Post-starbursts}

Many post-starburst galaxies show obvious tidal features or disturbed morphologies, providing evidence for a recent merger or interaction. A range of post-starburst morphologies can be seen in the SDSS images in Figure \ref{fig:manga}. \citet{Zabludoff1996} observed tidal features in $\sim$25\% of post-starburst galaxies selected based on their Balmer absorption H$\beta\gamma\delta$ strengths and lack of [OII] emission, using Digital Sky Survey imaging. \citet{Blake2004} observed evidence for interaction in $\sim10-20$\% of post-starburst galaxies selected in a similar way. \citet{Zabludoff1996} predicted a higher fraction of post-starburst galaxies would show such features in deeper imaging.  Indeed, studies using SDSS, Gemini, and {\it HST} imaging find such features in $\sim$50\% of post-starbursts. \citet{Pracy2009} observed tidal features or disturbed morphologies in 5/10 of the post-starburst galaxies imaged using Gemini/GMOS, using the same selection method as \citet{Zabludoff1996} and \citet{Blake2004}. \citet{Yang2004} and \citet{Yang2008} observed the sample of post-starburst galaxies from \citet{Zabludoff1996} using {\it HST}, finding tidal features or disturbed morphologies in 11/21 galaxies. \citet{Brown2009} observed tidal features in 10/24 post-starburst galaxies in the NOAO Deep Wide-Field Survey, selected based on strong H$\delta\gamma$ strengths and weak H$\alpha$ emission.

\citet{Pawlik2015} developed a new measure of galaxy Asymmetry to detect faint tidal features, and found 45\% of young post-starbursts (with $<100$ Myr since the starburst; from the PCA-selected sample of \citet{Wild2010}) to have tidal features. \citet{Pawlik2015} observed the fraction of post-starburst galaxies with observable tidal features to decrease with age, with $\sim$ 25-30\% of post-starbursts $>300$ Myr from the starburst having tidal features (for the same \citet{Wild2010} sample). The declining fraction of tidal features with post-starburst age is consistent with simulations of galaxy mergers \citep[e.g.,][]{Lotz2008}. 

\citet{Sazonova2021} considered a variety of metrics to measure the presence of disturbed morphologies in post-starburst galaxies, using \textit{HST} snapshot observations with higher spatial resolution yet similar depths as SDSS imaging, finding morphological disturbances in almost 90\% of the galaxies from the molecular gas rich sample of \citet{Alatalo2016b}. These disturbances were often in the inner regions of the galaxies, suggesting either mergers or secular processes could have been the cause. Similar to \citet{Pawlik2015}, \citet{Sazonova2021} find the presence of disturbed and asymmetric features to decline with time after a starburst, such that older post-starburst galaxies may no longer have observable merger features if they were once present.

Post-starburst galaxies are occasionally observed to have companion galaxies. \citet{Yamauchi2008} find the incidence of companions around the \citet{Goto2007} sample to be 8\%, higher than the 5\% companion incidence around normal galaxies. These galaxies may have experienced an initial starburst upon the first passage of the merger, and may experience another burst upon coalescence \citep[e.g.,][]{Mihos1994, Snyder2011}.

\subsection{Bulge vs. Disk components in Low Redshift Post-Starbursts}

Post-starburst galaxies are generally bulge-dominated, with a lack of disks, and \Sersic indices ranging from $n\sim2$ though $n>8$ depending on the resolution of the data, the method used, and the galaxy selection method. 

Studies using SDSS imaging have found bulge-dominated profiles, yet with \Sersic indices $n\sim2-3$, lower than the typical \dev early type galaxy profile. \citet{Quintero2004} selected a sample of 1194 galaxies using SDSS spectra, selecting on galaxies with model fits showing high A/K light components and a lack of H$\alpha$ emission. Using SDSS imaging, \citet{Quintero2004} observed the vast majority (90\%) to be bulge-dominated, with the remaining 10\% also showing significant bulge components. The post-starburst sample had intermediate \Sersic indices of $n\sim3$, in between early type and star forming galaxies, and high surface brightnesses, higher even than that of early type galaxies, consistent with evolving to early type levels once the young stellar populations fade. \citet{Mendel2013} selected post-starburst galaxies using a similar A/K-based selection method, cutting on D$_n$4000 to remove galaxies with current star formation, and used SDSS imaging, finding post-starburst galaxies to have \Sersic indices of $n>2$, noting a lack of true disks with $n<1.5$. \citet{Pawlik2015} selected post-starburst galaxies using the PCA method of \citet{Wild2010}, finding these galaxies to have \Sersic indices $n\sim1-3$, with many typical of early type disks.

Higher resolution {\it HST} imaging has shown higher \Sersic indices for post-starburst galaxies. \citet{Yang2004} observed five post-starburst galaxies from the \citet{Zabludoff1996} sample with {\it HST}, finding bulge-dominated morphologies with bulge to total light fractions B/T$>0.5$ and \Sersic indices $n\sim5-9$ higher than the \dev $n=4$ typical of early type galaxies. One galaxy in this sample was too disturbed to be well-modeled by smooth components, due to a clear on-going merger. Consistent with predictions that gas-rich galaxy mergers could create power law elliptical galaxies \citep{Lauer1995, Faber1997}, the post-starburst galaxy surface brightnesses were found to be similar to those of power-law ellipticals, with higher overall normalization due to the brighter young stellar populations. In a follow-up study of the entire \citet{Zabludoff1996} sample, \citet{Yang2008} found typical B/T ratios for poststarbursts to be B/T$\sim0.59$, and high \Sersic indices of $n>5$ in 17/20 galaxies. In the remaining three galaxies, \citet{Yang2008} observed dust obscuring the central region of the galaxy, which may have affected the fits. These high \Sersic indices appeared to be caused by a variety of central structures, from bright nuclei to bars and rings. \citet{Pracy2009} used Gemini GMOS imaging on a similarly-selected sample of post-starburst galaxies, finding 9/10 to be consistent with \dev profiles and one disk.

\subsection{Stellar Population Gradients in Low Redshift Post-starbursts} 
 
Post-starburst galaxies typically have centrally concentrated young stellar populations, that are nonetheless spread outside of the nuclear regions. Even without spatially-resolved spectroscopy, the concentrated young stellar populations in post-starburst galaxies can be observed via their blue color gradients, with bluer colors observed in at small radius than at large radius \citep{Yang2008}. Centrally concentrated young stellar populations are expected for starbursts triggered by a major merger \citep[e.g.][]{Mihos1994,Bekki2005, Hopkins2009}, with the relative distribution of the younger stellar population sensitive to the galaxy's recent merger history.
 
Post-starburst galaxies typically have strong gradients in their Balmer absorption lines, with stronger absorption near the center of the galaxy. This reflects the central concentration of young stars formed in the recent burst. Given the age-burst degeneracy (\S\ref{sec:stellarpops}), further modeling is needed to determine whether the H$\delta$ gradient is driven primarily by the centrally concentrated burst stellar populations with respect to the more extended older populations, or whether starbursts proceed outside-in.  Simulations by \citet{Zheng2020} find the former to be the case.  Spatially-resolved studies of post-starburst stellar populations have found varying characteristic sizes for the post-starburst signature. Some \citep{Caldwell1996, Norton2001,Swinbank2012} find the region of strong Balmer absorption to extend over a radius of $\sim2-3$ kpc, while others \citep{Pracy2012,Pracy2013} find the strong H$\delta$ absorption limited to the central $<1$ kpc. However, the redshift ranges, stellar masses, and selection methods for each study vary. \citet{Pracy2013} consider a post-starburst sample selected at the ``sweet spot" of $z\sim0.02-0.04$, where massive galaxies can be selected given the sufficient volume, and seeing-limited observations can still resolve at $<$kpc spatial scales, finding strong H$\delta$ gradients limited to the central kpc of each of the four post-starburst galaxies considered. In order to understand and study these trends, a systematic study of the stellar population gradients of large numbers of post-starburst galaxies is needed. A range of H$\delta$ gradients can be seen in the MaNGA maps shown in Figure \ref{fig:manga}.
 
Galaxies selected to have individual MaNGA spaxels with post-starburst signatures show most galaxies to have irregular distributions of post-starburst regions (292/360), with relatively few having post-starburst regions confined to their centers (31/360) or ring-like morphologies (37/360) \citep{Chen2019}. The MaNGA fibers trace regions $\sim1$ kpc at $z\sim0.03$, the typical redshift of the MaNGA sample. Selecting post-starburst galaxies using IFU spaxels often finds off-center regions in otherwise star-forming galaxies \citep{Rowlands2018}. Considering only the central or ring post-starburst morphologies, \citet{Chen2019} find the central post-starbursts to have low star formation over the entire galaxy, dispersion-dominated kinematics, and merger signatures in contrast to the ring post-starbursts which are often still star-forming in their centers. These galaxy-wide differences suggest central post-starbursts have been triggered by different mechanisms than ring post-starbursts, and that MaNGA-selected central post-starbursts have properties similar to post-starbursts selected using fiber or slit spectroscopy. Extended star-forming regions were found in the outskirts of post-starburst galaxies in \citet{Pracy2014a}, although these galaxies were at lower redshifts ($z<0.01$) than the others considered here. Although aperture bias is always a concern for galaxies selected from their central properties, most samples of post-starburst galaxies selected using their central properties do not show extended star-forming regions.

\subsection{Morphologies and Structure at High Redshift}

Post-starburst galaxies at redshifts $z\sim0.8-2.5$ are compact, consistent with the ``red nugget" compact quiescent galaxies observed around cosmic noon \citep{Whitaker2012}. When the sizes and stellar masses of post-starburst galaxies are considered, as in \citet{vanderwel2014}, post-starburst galaxies are observed to have distributions consistent with quiescent galaxies, with light-weighted sizes smaller than those of quiescent galaxies, and high \Sersic indices \citep{Yano2016, Almaini2017, Wu2018,Maltby2018}. Given the differing color gradients in post-starburst galaxies compared to quiescent galaxies, \citet{Suess2020} find that if mass-weighted sizes are used, the difference between post-starburst and quiescent galaxy sizes is removed, with both still significantly smaller than star-forming galaxies. These compact high redshift post-starburst galaxies are consistent with the centrally-concentrated star-formation from an episode of compaction \citep{Zolotov2015}, triggered possibly by a merger or gas-rich disk instability.

\citet{Maltby2018} observe differing trends in size and \Sersic index for galaxies $z<1$ and $z>1$, that are somewhat in conflict with the low redshift results discussed above. This may be due to differences in selection and stellar mass. The selection by \citet{Maltby2018} uses 8 photometric colors and a principal component analysis that identifies unique spectral shapes \citep{Wild2014,Wild2016, Maltby2016}. As discussed in \S\ref{sec:cosmictime}, post-starburst galaxies at high redshift are fundamentally different than galaxies evolving through the post-starburst phase at low redshift, in that the latter have a burst population on top of some existing stellar population, whereas high redshift post-starburst galaxies could have formed the majority of their stars in the recent starburst dominating their stellar light. If low redshift galaxies are ``K+A", high redshift galaxies can be simply ``A". This difference likely accounts for the discontinuity in redshift trends observed in higher redshift samples. \citet{Maltby2018, Suess2021} observe a lack of high mass post-starburst galaxies at lower redshifts. \citet{Maltby2018} observe \textit{no} post-starburst galaxies at log stellar mass $>10$ at $z<1$, in contrast to the many observed in the low redshift samples listed above, and even samples of A star-dominated high stellar mass galaxies at $z\sim0.7$ \citep{Setton2020}. It may be that only low stellar mass galaxies at low redshift have single-epoch stellar populations that continue the trends observed at high redshift, which would be consistent with the higher burst mass fractions observed in lower mass galaxies \citep{French2018}.

The stellar population gradients in high redshift post-starburst can be traced by patterns in H$\delta$ absorption, as done above for low redshift galaxies. Studies so far show differing results, from flat H$\delta$ gradients observed by \citet{Setton2020} at $z\sim0.6$ and flat color gradients observed by \citet{Suess2021} at $1<z<2.5$, to strong H$\delta$ gradients similar to those observed in low redshift post-starbursts by \citet{DEugenio2020} at $z\sim0.8$. The nature of the stellar population gradient in post-starburst galaxies at high redshift can shed light on the mechanisms that caused the starburst and it's subsequent end. Further modeling is needed to find the cause of the differing H$\delta$ gradient results. Given the age-burst degeneracy (\S\ref{sec:stellarpops}), the contribution from the underlying old stellar population can result in different H$\delta$ gradient trends between samples selected at different redshifts. Comparing the radial trends in starburst age (distinct from the light weighted stellar population age) will shed light on whether high and low redshift starbursts have similar spatial distributions, or whether differing starburst triggering mechanisms result in differing spatial distributions. Upcoming NIR IFU instruments on JWST and 30m class telescopes will yield important information for these galaxies.

Post-starburst galaxies in high redshift clusters show signs of different mass-size evolution than those in less dense environments. \citet{Matharu2019} find $z\sim1$ cluster post-starbursts to be in between star forming and quiescent galaxies on the size-mass relation, with post-starbursts having larger sizes than quiescent galaxies at a given stellar mass.

\section{Kinematics and Dynamics}
\label{sec:kin}

The kinematics of post-starburst galaxies provide important clues to their recent merger histories. Post-starburst galaxies tend to be dispersion-dominated, with $v/\sigma \sim 0.1-1$. An early study by \citet{Norton2001} used long-slit spectroscopy to study a sample of 20 post-starburst galaxies from \citet{Zabludoff1996}, selected based on their Balmer absorption H$\beta\gamma\delta$ strengths and lack of [OII] emission, with redshifts 0.05--0.12. \citet{Norton2001} found 18 of the 20 galaxies to have $v/\sigma >1$ (dispersion-dominated), and only 6/20 galaxies to have any measurable rotation with $v>40$ km/s. Using integral field spectroscopy, \citet{Pracy2009} found a similar range of $v/\sigma$ values in a sample of 8 post-starburst galaxies, though with a higher mean value of $v/\sigma$ and a higher fraction of galaxies with measurable rotation $v>40$ km/s, despite the similar selection method and redshift range of the sample. These observed differences may be caused by the small sample sizes or different measurement techniques; observations with large IFU surveys will yield important new information. 

The typical values of $v/\sigma \sim 0.1-1$ found by both \citet{Norton2001} and \citet{Pracy2009} can be compared to simulations of major mergers to connect the galaxies back to possible merger progenitors. One such example by \citet{Naab2003} finds that more equal mass mergers result in lower $v/\sigma$ remnants than unequal mass mergers; 1:1 mass ratio mergers typically result in a galaxy with $v/\sigma <0.4$; 2:1 in $v/\sigma \sim 0.2-0.8$, 3:1 in $v/\sigma \sim 0.4-0.9$, and 4:1 in $v/\sigma \sim 0.5-1.1$. The observed sample of post-starburst galaxies likely results from a range of these merger scenarios, with both major $\sim$1:1 as well as the more common unequal mass scenarios. The presence of more minor $\sim$3:1 merger remnants in post-starburst samples is consistent with hydrodynamic simulations including radiative transfer by \citet{Snyder2011}, who found the strong H$\delta$ post-starburst signature can last up to 200 Myr in the period after a 3:1 merger, although more equal mass merger remnants will produce longer-lived post-starburst signatures.

IFU observations allow additional signatures of the kinematics to be connected to merger histories. A range of post-starburst velocity maps and velocity dispersion maps can be seen in Figure \ref{fig:manga}. \citet{Emsellem2007} introduced the $\lambda_R$ measure of the projected stellar angular momentum per unit mass, finding early type galaxies to separate into ``slow" and ``fast" rotators. Using this framework, IFU observations of post-starburst galaxies show them to be primarily fast rotators \citep{Pracy2009,Swinbank2012,Pracy2013}. 
The post-starburst galaxies are similarly distributed in $\lambda_R$ and ellipticity as comparison early type galaxies, consistent with evolving into that population. While the lack of slow rotators among the post-starburst sample is sometimes interpreted as a lack of significant recent mergers, mass ratios even as close to equal as 2:1 will still result in fast rotators \citep{Emsellem2011}. The cases of slow rotator post-starbursts may represent the more rare scenario of a truly equal mass merger, although simulated slow rotators are often galaxies which formed early, and have experienced many Gyr of subsequent minor mergers \citep{Naab2014}. Larger samples and further study will be required to understand whether rapid evolution through the post-starburst phase can create the entire observed sample of slow rotating early types, or whether these galaxies must be affected by a series of multiple mergers. Given the competing effects of equal mass mergers being rare while producing a longer-lived post-starburst signature, a more detailed study of post-starburst progenitors could test whether various merger scenarios can effectively end star formation in a galaxy.

Kinematic studies provide an opportunity to test whether the mechanisms that produce post-starburst galaxies change with redshift. A study of one z$\sim0.7$ post-starburst galaxy by \citet{Hunt2018} measured a $v/\sigma > 0.3$, possibly consistent with the low redshift samples. The galaxy showed measurable rotational support, although the value of $v/\sigma$ is a lower limit due to the effect of beam smearing. With larger samples of measurements like these, the possibility of different post-starburst channels at different redshifts can be explored.

\section{The interstellar medium}
\label{sec:ism}

\begin{figure}
    \centering
    \includegraphics[width=0.5\textwidth]{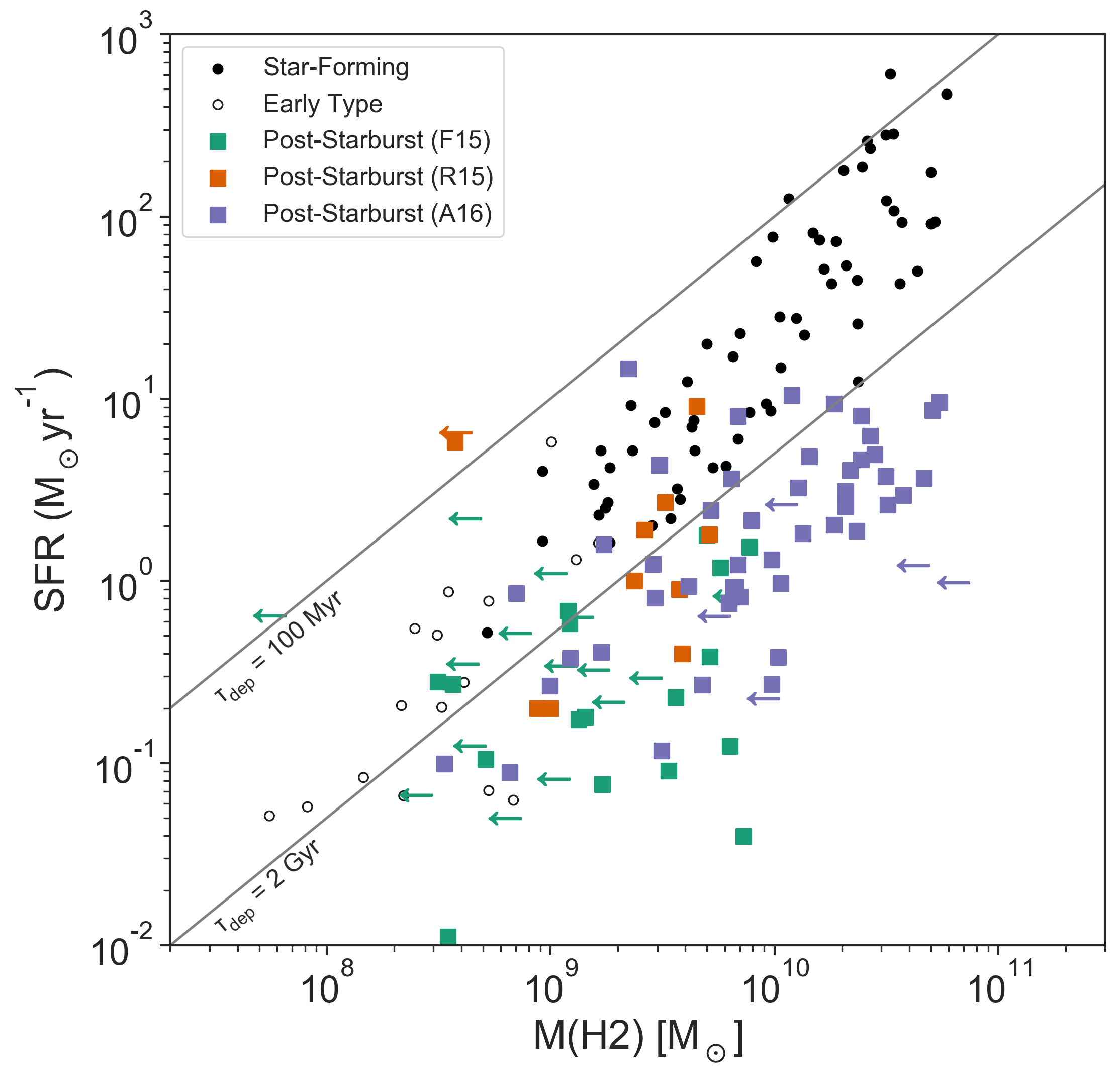}
    \caption{Molecular gas masses vs. SFR for post-starburst galaxies and comparison samples. Post-starburst samples are shown from \citet{French2015,Rowlands2015,Alatalo2016b} with SFRs measured using H$\alpha$ luminosities, using the Balmer decrement to correct for dust, and using the emission line ratio method from \citet{Wild2010} to correct for AGN contributions to the H$\alpha$ luminosity. All M(H$_2$) measurements are from CO (1--0) line measurements, and all use the same value of the CO-to-H$_2$ conversion factor $\alpha_{CO} =4$. Comparison samples are shown from \citet{Young2011} (early type galaxies) and \citet{Gao2004} (star-forming and starbursting galaxies). The post-starburst galaxies have higher molecular gas masses than would be expected from their low SFRs. Depletion times of 100 Myr and 2 Gyr are indicated with grey lines. Most of the comparison galaxies have molecular gas depletion times within this range, whereas most of the post-starburst galaxies have longer depletion times.
    }
    \label{fig:gas}
\end{figure}

The state of the interstellar medium in post-starburst galaxies is of particular importance to understanding the physical processes that drive the rapid decline in star formation. Despite the low SFRs in post-starburst galaxies, significant amounts of atomic and molecular gas, dust, and PAHs have been observed in a variety of post-starburst galaxy samples.

\citet{Chang2001} surveyed the HI content of a sample of post-starburst galaxies, detecting atomic gas in 1/5 galaxies targeted from the \citet{Zabludoff1996} sample. In contrast, \citet{Bravo?Alfaro2001} observed a lack of HI in 11 post-starbursts in a dense cluster environment, suggesting cluster specific processes may affect the gas during this phase in a different way. Subsequent observations have found post-starburst galaxies to have atomic gas fractions in between those of early and late type galaxies; \citet{Buyle2006} detected HI in 4/6 post-starbursts and \citet{Zwaan2013} in 6/11 post-starbursts selected from the catalogs of \citet{Zabludoff1996, Goto2003, Goto2007}. A spatially-resolved follow-up study of the system detected in HI by \citet{Chang2001} by \citet{Buyle2008} revealed the HI to be distributed around the two merging galaxies in tidal streams, with a lack of HI directly coincident with the galaxies, suggesting the atomic gas supply had been disrupted by the merger.

Cold molecular gas traced by CO has been observed in several nearby post-starburst galaxies, NGC 7252 \citep{Dupraz1990} and NGC 5195 \citet{Kohno2002}. Studies of post-starburst galaxies selected using H$\delta$ absorption and a lack of H$\alpha$ emission \citep[33 galaxies]{French2015}, the PCA selection method \citep[11 galaxies]{Rowlands2015}, and shocked post-starbursts \citep[52 galaxies]{Alatalo2016b} have all observed molecular gas fractions in post-starburst galaxies more similar to star forming galaxies than quiescent early types. Similarly high molecular gas fractions are inferred by \citet{Yesuf2020} using Balmer decrement H$\alpha$/H$\beta$ measurements. 

These observations of cold molecular gas in post-starburst galaxies indicate that starbursts can end without the complete removal or consumption of their molecular gas reservoirs. The star formation efficiencies in the remaining gas are lower than normal galaxies, with the post-starbursts typically lying below the Kennicutt-Schmidt relation \citep{Kennicutt1998}. The molecular gas masses and star formation rates of these samples, in comparison to other galaxy types, are shown in Figure \ref{fig:gas}. 

\citet{Suess2017} found higher molecular gas masses in a sample of two z$\sim0.7$ post-starbursts than would be expected given their low SFRs, consistent with the lower redshift samples. \citet{Belli2021} find similarly high molecular gas fractions ($\sim10-20$\%) in a sample of three massive quiescent galaxies at $z\sim1-1.25$, including one post-starburst. Other studies at higher redshift have observed a lack of molecular gas in recently quenched massive galaxies; \citet{Williams2021} find gas fractions $<$6\% in a sample of 6 $z\sim1.5$ quiescent galaxies and \citet{Morishita2021} find a gas fraction $<5$\% in a $z=1.91$ massive quiescent galaxy. 

Observations of denser molecular gas traced by HCN and HCO$+$ have found low dense-gas fractions in post-starburst galaxies \citep{French2018b}, consistent with the low current SFRs, and which would explain their low CO-traced star formation efficiencies if something is preventing the CO-traced gas from collapse.
However, different behavior is observed for the more nearby systems NGC 5195 \citep{Kohno2002,Matsushita2010,Alatalo2016a} and NGC 1266 \citep{Alatalo2015}. Observations of NGC 5195 are complicated by its on-going merger with M51, but is observed to have a star formation efficiency similar to 
normal galaxies. NGC1266 has a high HCN/CO ratio, which may be due to an embedded AGN \citep{Juneau2009}. Observations of the dense gas in larger samples of post-starburst galaxies are needed to understand how and when galaxies might lose their dense gas reservoirs.

Large dust reservoirs are seen in post-starburst galaxies, traced by IR photometry from \textit{WISE}, \textit{Spitzer}, and \textit{Herschel}. \citet{Rowlands2015} observed dust masses similar to those in star-forming galaxies in a sample of 11 PCA-selected post-starbursts, with dust mass and dust temperature declining with starburst age. \citet{Smercina2018} observed compact, warm dust reservoirs in 33 H$\alpha$-H$\delta$ selected post-starbursts, and suggested the unique A-star dominated stellar population leads to a ``high-soft" radiation field. \citet{Li2019} used archival photometry to measure dust masses for a sample of 58 post-starburst galaxies selected using a variety of methods, observing a trend of decreasing dust mass to stellar mass fractions with post-starburst age consistent with the trend observed for CO-traced molecular gas. \citet{Alatalo2016c} observed an excess of 22$\mu$m flux in a sample of post-starburst galaxies from \citet{Goto2007}, indicating the possible influence of AGN in heating the dust.

PAHs were observed by \citet{Roseboom2009} in a sample of 11 post-starbursts observed with \textit{Spitzer}. Several PAH ratios were observed to evolve with post-burst age, consistent with low level AGN activity. \citet{Smercina2018} observed high PAH abundances in a sample of 33 post-starbursts, with PAH/TIR fractions higher than typical star-forming galaxies.

The circumgalactic medium (CGM) contains clues to past outflows from galaxies and the future possibility of gas accretion. QSO line of sight observations of a post-starburst galaxy by \citet{Tripp2011} show a large reservoir of ``warm-hot" gas at $10^{5.5}$ K and evidence for a multiphase wind. \citet{Heckman2017} observed the absorption of Ly$\alpha$, Si [III], C [IV], and O [VI] in a sample of 17 galaxies from the \citet{Wild2007} sample of PCA-selected post-starbursts with background QSOs. \citet{Heckman2017} find stronger absorption out to larger radii in the post-starbursts and at higher velocities than in comparison star-forming galaxies, with evidence for a starburst-driven wind interacting with pre-existing gas. At higher redshifts, the CGM is more difficult to constrain. \citet{Zahedy2020} observe a transient warm (T$\sim10^5$ K) phase in the CGM around a $z\sim0.4$ massive quiescent galaxy, consistent with heating from AGN activity or from evolved stellar populations, which provide a clue to how these galaxies might stay quiescent.

Combining the molecular gas measurements of \citet{French2015,Rowlands2015,Alatalo2016b}, \citet{French2018} find a trend of declining molecular gas fraction with post-starburst age. This trend has a characteristic timescale of $\sim 200$ Myr, and galaxies would reach the low molecular gas fraction in early type galaxies \citep{Young2011} in $1-2$ Gyr. This trend is consistent with the timescale for the dust mass fraction to decline during the post-starburst phase \citep{Li2019} and the timescale for the star-forming gas in simulated post-starburst galaxies to decline in the EAGLE simulation \citep{Davis2018}. These timescales are much shorter than the gas depletion timescales from the low levels of on-going star formation, suggesting AGN feedback may be required to remove the gas and dust on these short timescales.

\section{AGN and Nuclear activity}
\label{sec:agn}

\begin{figure*}
    \centering
    \includegraphics[width=0.32\textwidth]{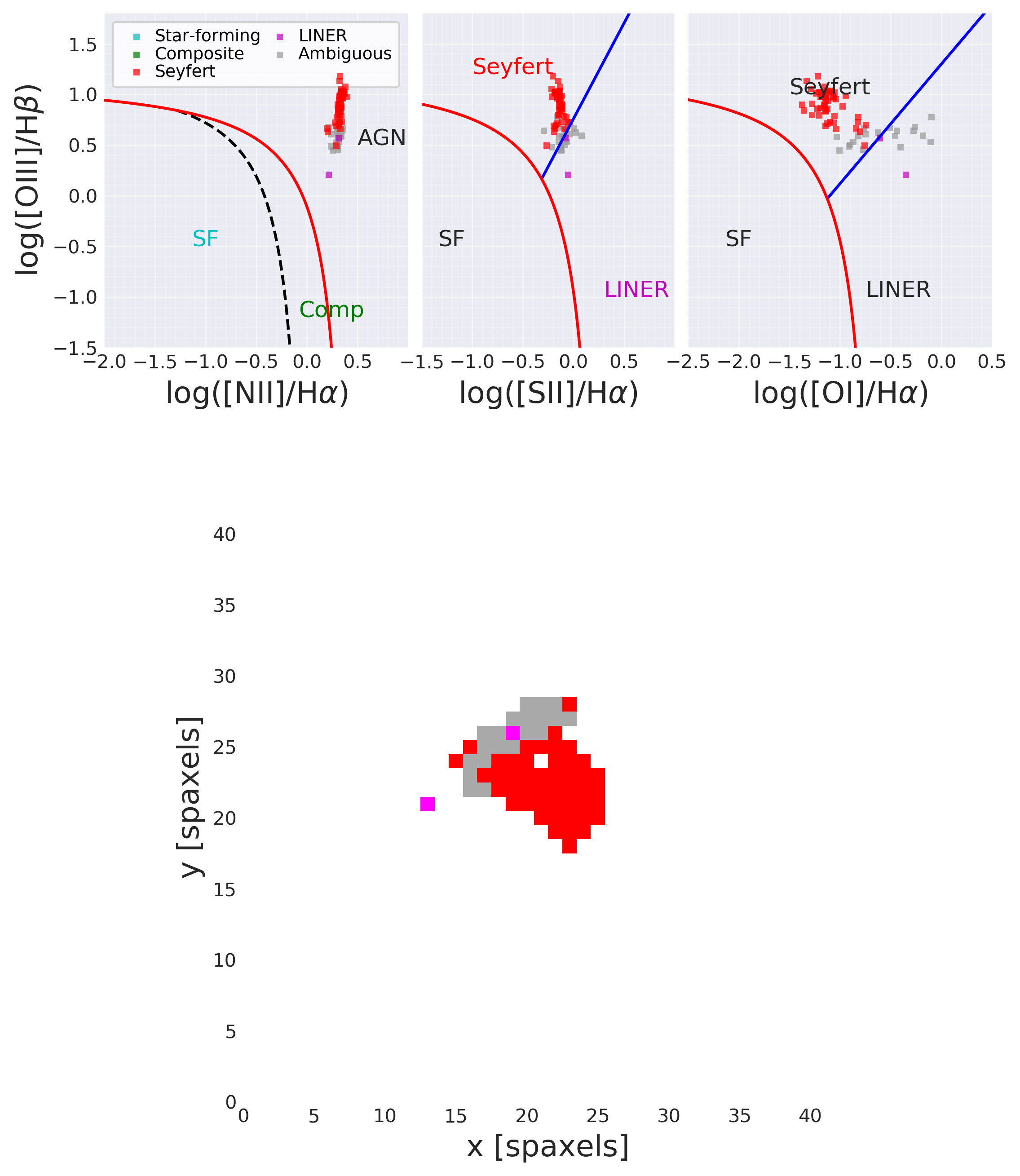}
    \includegraphics[width=0.32\textwidth]{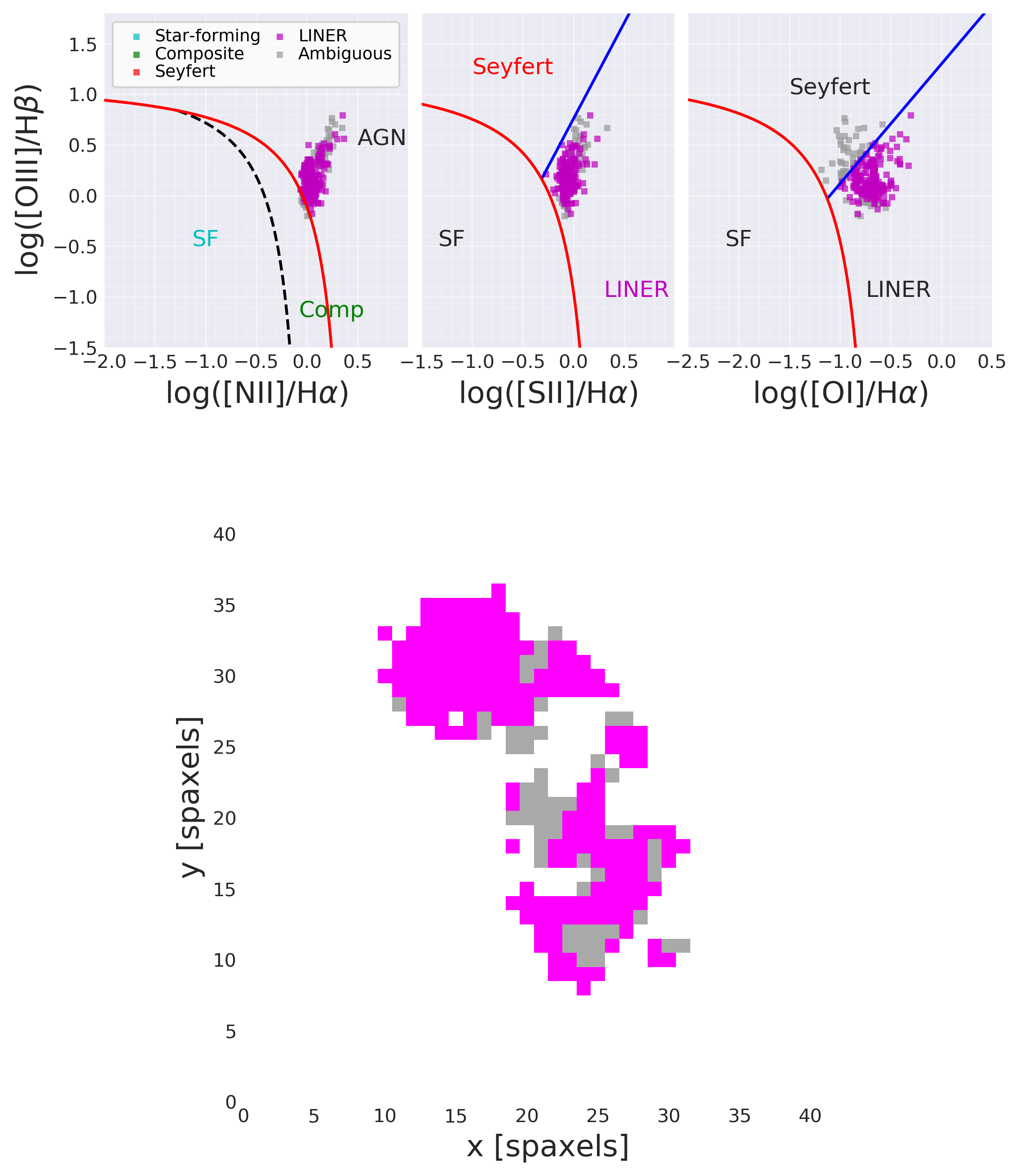}
    \includegraphics[width=0.32\textwidth]{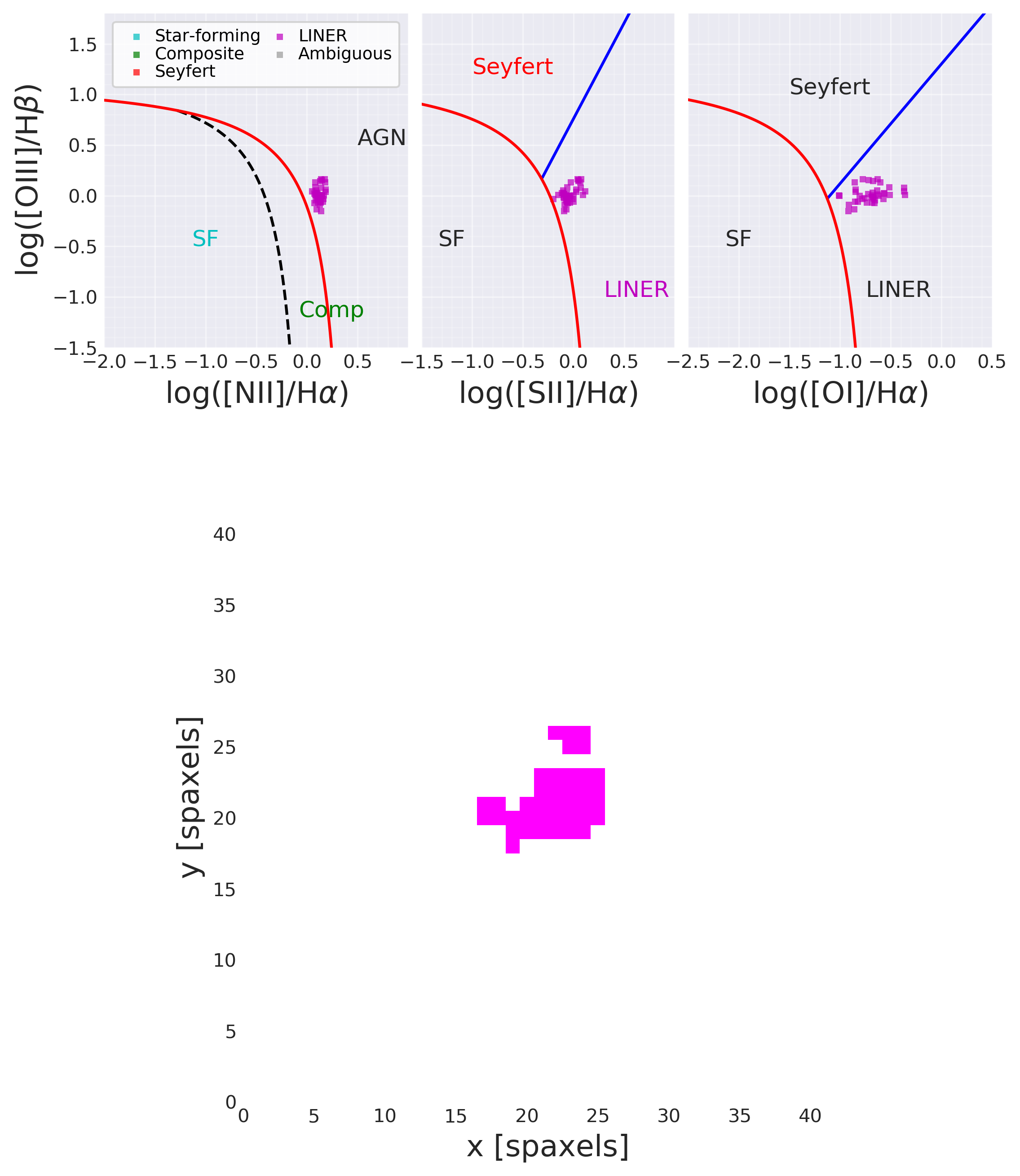}
    \includegraphics[width=0.32\textwidth]{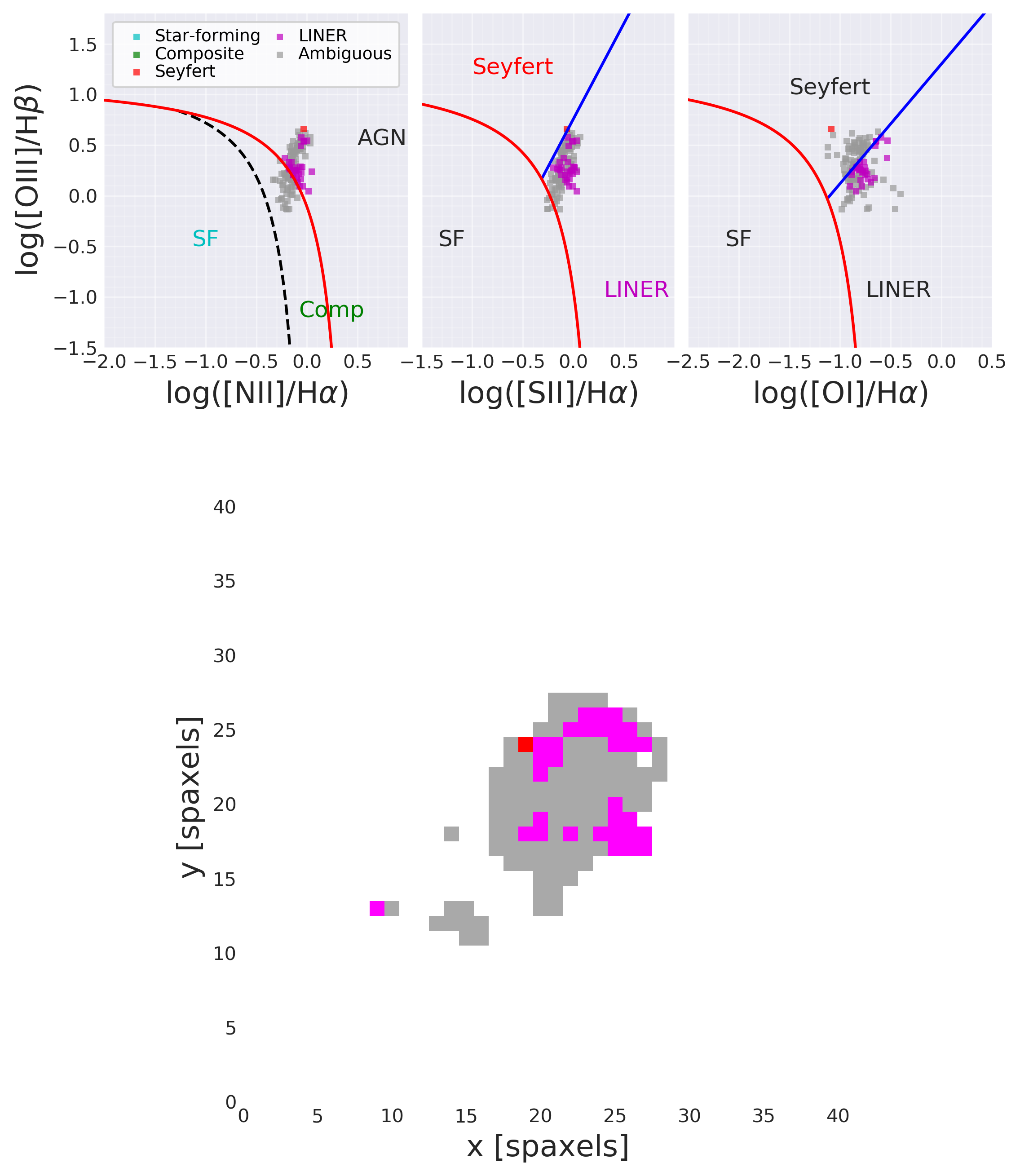}
    \includegraphics[width=0.32\textwidth]{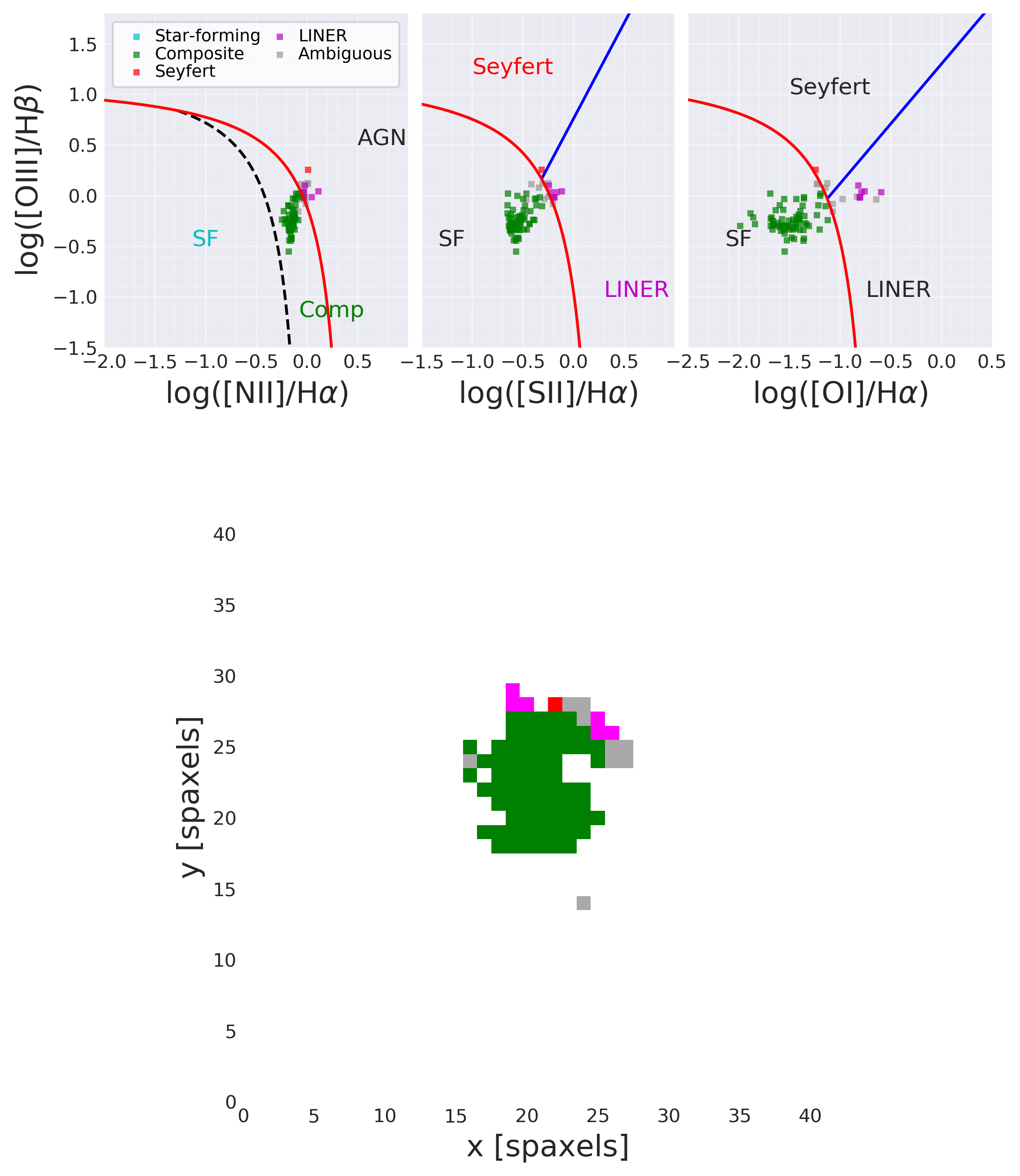}
    \includegraphics[width=0.32\textwidth]{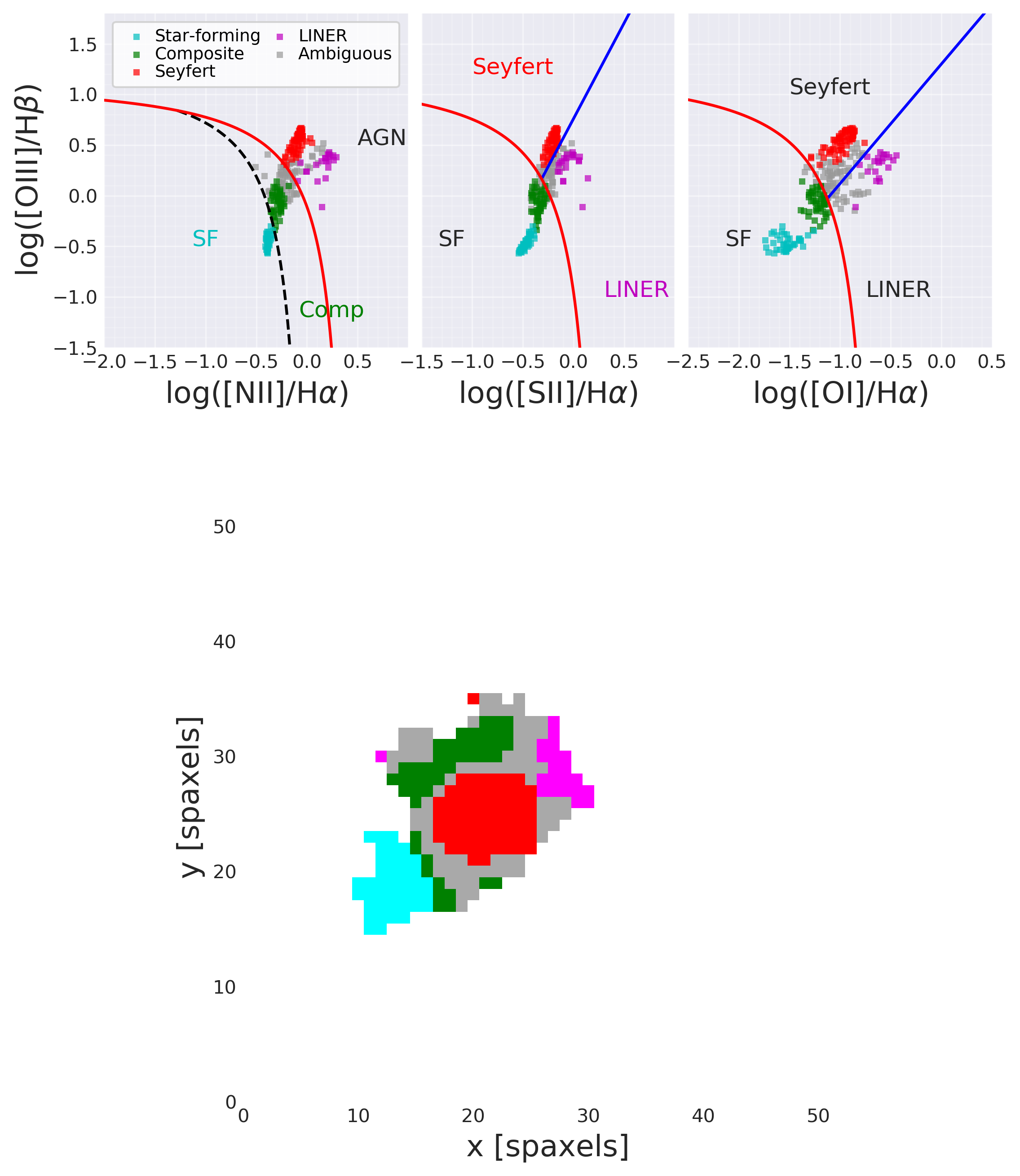}
    \caption{Emission line ratio diagnostics and color-coded (star-forming in cyan, composite in green, Seyfert in red, LINER in pink, and ambiguous in grey) spaxel maps from MaNGA observations \citep{manga,marvin} of the same sample of post-starburst galaxies as in Figure \ref{fig:manga}. These galaxies are in the same order from the top left (top: 9494-3701, 8555-3701, 8997-3703; bottom: 8080-3702, 9194-3702, 8329-6102). The galaxies show a range of emission line ratio patterns, from concentrated Seyfert emission in the center, to distributed LINER-like or composite emission. 
    }
    \label{fig:bpt}
\end{figure*}

Observing AGN activity in post-starburst galaxies is complicated by the need to select against on-going star-formation, variability of AGN, and the variety of AGN signatures. Despite these challenges, understanding the presence and impact of AGN in these systems is of vital importance to models of AGN feedback. Delays in time between the starburst/merger and the onset of AGN activity are expected from theory due to the angular momentum loss required in the gas to reach the nucleus \citep{Hopkins2012b}, and AGN feedback may require an initial disruption of the star-forming disk to be effective \citep{Pontzen2017}.

Optical emission line ratio diagrams \citep{Baldwin1981, Kauffmann2003} are sensitive to the presence of Seyfert II AGN and low luminosity AGN displaying LINER emission. Many studies have found post-starburst galaxies to have emission line ratios characteristic of LINERs  \citep{Caldwell1996,Yan2006,Yang2006,Wild2007,Brown2009, Wild2010,Mendel2013, DePropris2014, French2015,Alatalo2016}\footnote{Many of the younger post-starbursts in selection methods that do not select against on-going star formation have BPT line ratios more consistent with star formation \citep[e.g.][]{Wild2007,Wild2010}}. Studies using unresolved spectroscopy are limited in their interpretation, especially of LINERs. The nature of LINERs is ambiguous, especially during the post-starburst phase. LINER emission line ratios can also be generated by shocks \citep[e.g.,][]{Rich2015} or post-AGB stars \citep{Yan2012}. These alternate possibilities are especially likely when the emission line strengths are weak, and the galaxies are ``LINER-like" on a WHAN diagram \citep{CidFernandes2011}. Spatially resolved spectroscopy is required to distinguish between distributed non-nuclear LINER signatures (``LIERs", \citealt{Belfiore2016}) from galaxies with nuclear emission from AGN. Several examples of spatially resolved spectroscopy from MaNGA \citep{manga,marvin} are shown in Figure \ref{fig:bpt}; the post-starburst galaxies show a range of emission line ratio patterns, from concentrated Seyfert emission in the center, to distributed LINER-like or composite emission. Combining spatially and kinematically resolved observations can be especially powerful in determining the role of high velocity dispersion shocked components \citep{Rich2015,Davies2016, DAgostino2019, Law2020}. 

\citet{Wild2010} used the location of galaxies on the BPT diagram to disentangle emission from star formation from black hole accretion, finding the black hole accretion rate to peak well after the starburst, 250 Myr into the post-starburst phase. \citet{Greene2020} find evidence for AGN in $z\sim0.7$ post-starburst galaxies using both optical emission line ratios and radio emission. The fraction of optically-selected AGN is observed to decline with post-starburst age, in contrast to the delay seen at lower redshift \citep{Wild2010}. However, we note that the definition of post-starburst age and the metric for AGN importance vary between these studies.

X-ray bright AGN have been observed in some samples of post-starburst galaxies. \citet{Brown2009} observed X-ray 0.5--7 keV luminosities of $\sim10^{41}$ erg/s in about a third of the post-starburst galaxies considered (at redshifts $\sim$0.1--0.3. At higher redshifts $\sim0.8$, \citet{Georgakakis2008} observed similar X-ray luminosities in a stack of {\it Chandra} observations of post-starburst galaxies. In contrast, \citet{DePropris2014} found {\it Chandra} 0.5--7 keV non-detections at $L_X < 10^{40}$ erg/s for a sample of 10 post-starburst galaxies at $z \sim 0.01-0.05$. This discrepancy may be related to the differing sample selection methods, especially if dust affects the selection. 

A small fraction of post-starburst galaxies show signs of weak, dust-obscured AGN activity. Obscured AGN can be identified via their hot dust signatures in the infrared. Post-starburst galaxies do not typically show evidence of obscured AGN via the WISE [3.4]-[4.6] color index \citep{Stern2012}; \citet{Alatalo2016c} find no post-starburst galaxies with WISE [3.4]-[4.6] $>0.8$. However, the composite SED studied by \citet{Alatalo2016c} showed a rise from 12 to 22 $\mu$m that could indicate the presence of buried AGN. In a sample of 33 low-redshift post-starbursts observed with {\it Spitzer} and {\it Herschel}, \citet{Smercina2018} find one post-starburst galaxy with signatures of a dust obscured AGN, with extremely red WISE color of [3.4]-[4.6]$=1.8$. In a sample of post-starburst galaxies selected using self-organized maps (see \S\ref{identifying}), \citet{Meusinger2016} find 3\% of post-starbursts with evidence for obscured AGN via their WISE colors [3.4]-[4.6] $>0.8$. Similarly low fractions of post-starburst galaxies show evidence of obscured AGN via radio emission. The FIRST survey \citep{first} at 1.4 GHz can constrain the presence of radio-loud AGN in these systems. \citet{Nielsen2012} found 4\% of post-starburst galaxies considered to have FIRST detections above the 5$\sigma$ noise level; \citet{Meusinger2016} find 2\% of their sample to have FIRST detections, at levels consistent with AGN activity instead of star formation. \citet{Smercina2018} find $\sim10$\% of their sample to have evidence for a possible radio excess.

Samples that are selected using cuts against emission lines find AGN with lower accretion rates than those without such cuts. The \citet{Zabludoff1996} sample was found to have low $L/L_{edd} \lesssim 0.1$\% \citep{DePropris2014} from a lack of X-ray emission and low [OIII] luminosities. In contrast, \citet{Wild2010} find many post-starburst galaxies selected using a PCA method to have higher accretion rates $L/L_{edd} > 1$\% using [OIII] luminosity measurements.

Many studies have focused on the class of quasars with post-starburst signatures (post-starburst quasars, PSQs). PSQs tend to fall into two categories: (1) spiral galaxies with weaker AGN and some on-going star formation (2) elliptical galaxies with more luminous AGN, signs of recent mergers, and more prominent burst populations \citep{Cales2013,Cales2015}. \citet{Cales2015} compared the stellar populations of PSQs and post-starbursts, finding younger burst ages in the post-starburst sample, implying the PSQ phase may occur during or after the quiescent post-starburst phase, instead of before it. However, these samples have differing redshift ranges, and comparisons among different selection methods can lead to systematic differences (see above in \S\ref{identifying}).

The timescale for AGN to vary from quasar-like to LINER-like is much faster than the timescale of the post-starburst phase \citep{Lintott2009, Keel2012}. One method to determine the typical AGN activity during the post-starburst phase is to compare the populations of galaxies with recent starbursts both with and without AGN optical line signatures. \citet{Pawlik2018} use this method to estimate the AGN duty cycle to be $\sim$50\% during the post-starburst phase. Another method is to use extended ionized regions to probe the AGN histories of individual galaxies. This method is insensitive to the details of the post-starburst selection method and allows for the AGN histories of unambiguously quiescent galaxies to be determined, but is significantly more observationally expensive. When galaxies are surrounded by clumps of gas, as post-starburst galaxies are likely to be after a merger, the gas can be illuminated by past AGN activity, tens of thousands of light years away from the nucleus of the galaxy. These extended emission line regions have been found in integral field spectroscopy, but extreme examples have also been found in broad band imaging. The most dramatic such case is Hanny's voorwerp \citep{Lintott2009}, where OIII$\lambda$5007 emission from an AGN light echo around a normal galaxy is bright enough to dominate the SDSS $g$ band. Follow-up work has revealed more such cases of ``voorwerpjes" \citep{Keel2012}, observed in SDSS imaging of AGN. These galaxies have past ionizing luminosities ranging from Seyfert-like to quasar-like. AGN light echoes have been observed in several post-starburst galaxies observed with integral field spectroscopy or narrow band imaging. MUSE observations \citep{Prieto2016} of a post-starburst galaxy\footnote{This galaxy was observed because it is the host galaxy of a Tidal Disruption Event, but the TDE is too recent to have ionized the extended emission line region.} revealed an OIII-bright AGN light echo.  
The ionized features extend 23$^{\prime\prime}$ (10 kpc at $z\sim0.02$), or $3\times10^4$ lightyears away from the nucleus, indicating they were ionized by AGN activity $\sim3\times10^4$ years ago. Similar features have been found in narrow band imaging in the post-starburst galaxies NGC 7252 by \citet{Schweizer2013} and M51b by \citet{Watkins2018}.

The accretion of individual stars is observed to increase during the post-starburst phase. These events are observed as Tidal Disruption Events (TDEs); when stars pass within the tidal radius of a supermassive black hole, tidal forces can overcome the self-gravity of the star. Much of the stellar debris will be accreted, releasing a bright flare. The TDE rate is observed to be higher in post-starburst galaxies than in other types of galaxies, with a rate increase high enough that many TDEs have been observed in post-starburst galaxies despite their rarity \citep{Arcavi2014,French2016,French2017,Law-Smith2017,Graur2018,French2020}. This effect cannot be driven by selection effects \citep[see][]{Roth2020}, but dust obscuration in starburst galaxies is likely limiting the observability of a high TDE rate during the starburst phase. The high TDE rate in post-starburst galaxies may be caused by the high central stellar densities during this phase \citep{Law-Smith2017,Graur2018,French2020b}, or unusual stellar dynamics \citep{Stone2018}. A more thorough discussion of these issues can be found in \citet{French2020}. The total mass growth of the supermassive black hole due to stars is likely less than the mass growth due to gas in the post-starburst phase. If a high TDE rate of $10^{-3}$ per year per galaxy persists over a Gyr, $\sim10^6$ M$_\odot$ would be accreted in the form of stellar debris. If 1\% of the $\sim10^9$ M$_\odot$ gas reservoirs are accreted by the supermassive black hole during this phase (an accretion rate of 1 M$_\odot$ yr$^{-1}$), $\sim10^7$ M$_\odot$ would be accreted from gas. On the other hand, the TDE rate is not subject to the same negative feedback effects as gas accretion, and the evolution of the TDE rate with time through this phase is still poorly constrained.

\section{The past: Possible causes for the starbursts and their ends}

\subsection{Mergers}

The presence of disturbed morphologies and tidal features, and the concentrated young stellar populations seen in many post-starburst galaxies are consistent with expectations for merger-triggered starbursts. Simulations of mergers between two gas-rich galaxies have shown this to be a pathway for galaxies to rapidly evolve through a starburst phase before becoming an early type galaxy \citep[e.g.,][]{Barnes1988, Barnes1991, Barnes1996}. The observed rates of post-starburst galaxies are consistent with merger rates as they evolve from the local universe out to at least $z\sim1$ \citep{Snyder2011}.

Simulations by \citet{Bekki2005} show that a variety of merger scenarios as well as strong tidal interactions can reproduce the observed population of post-starburst galaxies, with gas-rich major $\sim$1:1 mergers resulting in dispersion-dominated ellipticals, and more minor $\sim$3:1 mergers resulting in S0 type galaxies with more rotation. The range in observed kinematics in various post-starburst samples, discussed in \S\ref{sec:kin} is consistent with both major and minor mergers contributing to the post-starburst population. While minor mergers are more common than major mergers, the lifetime of the post-starburst signature is longer in simulated major merger remnants \citep{Snyder2011}, which will result in a disproportionate number of post-major-merger galaxies depending on the details of the post-starburst selection criterion. \citet{Davis2018} find major mergers $\sim1-3$:1 important in low redshift post-starburst identified in the EAGLE simulation, but that even micro mergers $\sim100:1$ are common in post-starburst progenitors, especially at higher redshifts. In contrast, \citet{Zheng2020} find that only major mergers can produce post-starburst galaxies, with enough star formation lingering in post minor merger galaxies to prohibit a post-starburst classification. This difference is likely driven by differing post-starburst selection criteria between \citet{Zheng2020} and \citet{Davis2018}; \citet{Zheng2020} impose a harsher cut on residual star formation, which selects for simulated galaxies with very little gas remaining after a merger-triggered episode of AGN feedback. As star formation and AGN feedback prescriptions evolve in simulations, the relative importance of each mechanism may shift.

Not all mergers will produce a starburst or a SFH that declines sharply enough to generate a post-starburst signature. \citet{Bekki2005}, \citet{Wild2009}, and \citet{Snyder2011} consider the production of post-starburst galaxies from simulations of merging galaxies spanning a range in parameters. Mergers between two gas-poor ($f_{gas} < 0.05$) galaxies will not result in a starburst, but if the gas fraction is too high ($f_{gas}\gtrsim40$\%), simulated starbursts have more prolonged declines that don't produce stellar populations that meet common post-starburst selection methods. \citet{Pawlik2019} consider several case studies of simulated EAGLE galaxies that pass through the post-starburst phase, finding one case where two gas-rich galaxies merge, and one case where a gas-rich galaxy triggers a starburst upon merging with a larger gas-poor galaxy. The bulge to disk ratios, relative orientations, and merger geometry will also affect the strength of the starburst and whether a merger remnant passes through a post-starburst phase \citep{Wild2009, Snyder2011}. This wide range in outcomes is likely the reason why some $\sim$3:1 mergers can result in a quiescent galaxy, others, like the merger between M31 and the progenitor of M32 \citep{DSouza2018} do not.

\subsection{Star Formation and Stellar Feedback}

One contributing factor to the end of a starburst is the consumption of molecular gas fuel and the disruption of giant molecular clouds (GMCs) by stellar feedback. The typical mass fractions of stellar mass produced during the starburst are a sizable fraction of the current stellar mass (see \S\ref{sec:stellarpops}), larger than or similar to the remaining molecular gas fractions ($\sim5-30$\%, see \S\ref{sec:ism}). Stellar winds are capable of driving outflows with mass loss rates exceeding the SFR \citep[e.g.,][]{Hopkins2012,Bolatto2013}.  The role of stellar feedback in ending star formation in massive galaxies is unclear: in some cases there is not enough energy to permanently eject the molecular gas supplies of more massive galaxies M$_* \gtrsim 10^{10.5}$ M$_\odot$ \citep[e.g.,][]{Veilleux2005}, while in other cases strong molecular outflows are observed in massive galaxies with compact starbursts that resemble the likely progenitors of post-starburst galaxies \citep{Geach2014, Geach2018}.

Studies of the post-starburst phase allow for a more highly time-resolved view of this picture. Does stellar feedback play a role in temporarily halting star formation at the beginning of the post-starburst phase? An epoch of star formation that is ended by its own stellar feedback is said to ``self-quench", and would have a short timescale of the order 10 Myr \citep{Schaerer1999,Stinson2007}, similar to the lifetime of a GMC \citep[see e.g.,][]{Kruijssen2019}. \citet{Wild2009} find that the initial end of the starburst in simulated post-starburst galaxies can indeed be due to gas consumption. \citet{McQuinn2010} test whether a sample of low mass galaxies have short starburst timescales consistent with ending due to gas consumption and stellar feedback, finding longer burst timescales $\sim400-1000$ Myr, and challenging the extremely bursty SFHs found in many simulations of dwarf galaxies. For more massive post-starburst galaxies, short burst durations $\lesssim50$ Myr are found for majorities of the low-redshift samples in \citet{Kaviraj2007a, French2018}, consistent with a few times that of the typical GMC lifetime. For the shorter burst-duration post-starbursts, stellar feedback may have some initial large role in shutting down star formation. However, characterizing SFHs down to 10 Myr precision is difficult in practice with integrated light measurements, so caution should be exercised\footnote{Especially when considering the very early post-starburst phase, the longer evolutionary timescales of massive stars predicted by binary evolution models \citep[and references therein]{Eldridge2020} will become increasingly important.}. Longer burst durations are commonly found for higher redshift poststarbursts, $\sim100-300$ Myr \citep{Wild2020,Forrest2020}, although this comparison is complicated by the differing sample selections and different SFH parameterizations used. Furthermore, if stellar feedback takes longer, $\sim100$ Myr to affect a galaxy through cumulative effects, it will be even more difficult to constrain the importance of this mechanism.

If gas consumption and stellar feedback play a role in the initial shutdown of star formation, AGN feedback (see next section) may be required to keep SFRs low during the post-starburst phase and prevent GMCs from re-forming \citep{Wild2009, Hopkins2013}.

\subsection{AGN feedback}

AGN feedback is thought to play a large role in driving galaxies from starbursting to quiescent, and there is growing evidence for how AGN feedback may operate. Observations of significant gas and dust remaining during the post-starburst phase (see \S\ref{sec:ism}) and on-going AGN activity (see \S\ref{sec:agn}) demonstrate that the way galaxies stop forming stars is more complex than a single phase of quasar gas blowout near the end of the starburst \citep[e.g.,][]{Hopkins2006}. 

Simulations can constrain the type of AGN feedback that may be acting during this phase. \citet{Zheng2020} considered several implementations of AGN feedback, finding mechanical feedback was necessary to explain the observed populations of post-starbursts. In these models, mechanical feedback effectively removed gas from galaxies to end star formation during the post-starburst phase, while thermal feedback models were less efficient. \citet{Davis2018} selected post-starburst galaxies from the EAGLE simulation, and found that while AGN activity was common during the post-starburst phase, strong outbursts were rare. AGN feedback may not require extremely high black hole accretion rates; \citet{Pontzen2017} used simulations of mergers to study the role of AGN feedback, and found that feedback can occur with low black hole accretion rates, much less than Eddington. \citet{Pontzen2017} found that black hole accretion can continue during the post-starburst phase, even while star formation is low, and that the disruption of the disk during the merger helps to make AGN feedback effective in shutting down star formation.

While dramatic molecular and multiphase outflows have been observed in quasars and other AGN samples \citep[e.g.,][]{Feruglio2010,Cicone2014}, these high molecular outflow rates have not been observed in post-starburst samples. \citet{Alatalo2011} observed a molecular outflow in the post-starburst galaxy NGC 1266, which is rapidly ($<100$ Myr) depleting the gas in the nuclear region of this galaxy. However, much of this gas will not be able to escape the galaxy \citep{Alatalo2015}, with an escape rate only $\sim2$ M$_\odot$ yr$^{-1}$.

Evidence for outflows in optical spectroscopy have been observed for a number of post-starburst galaxies identified via an outlier detection algorithm and selected to have strong H$\delta$ absorption \citep{Baron2017,Baron2018,Baron2020}.  These systems have large outflow rates, which may be able to remove the entire gas reservoir in a single episode of AGN activity. However, these systems each have significant on-going star formation, and may represent either an earlier phase of evolution to the post-starburst galaxies considered by other selection methods, or a separate track of galaxy evolution.

\citet{Baron2021} measured the presence of neutral gas outflows in a sample of galaxies selected to have strong H$\delta$ absorption and AGN-like emission line ratios as well as galaxies with evidence for ionized gas outflows like the sample discussed in the previous paragraph. 40\% of such galaxies showed neutral outflows, but with outflow rates smaller than those seen in normal AGN or ULIRGs.

Depending on the sample, starburst-driven and AGN-driven outflows can be difficult to distinguish. \citet{Tremonti2007} observed fast outflows $>1000$ km/s in Mg II $\lambda\lambda2796,2803$ absorption lines in a sample of $z\sim0.6$ galaxies with recent declines in their star formation histories, but follow-up observations found these galaxies to have significant levels of compact star formation, with current SFRs far above those of other samples of post-starbursts and capable of driving the observed outflows \citep{Sell2014}. Fe II, Mg II, and Mg I absorption in a sample of $z\sim0.2-0.8$ post-starbursts were measured by \citet{Coil2011} to have weaker outflows than those observed by \citet{Tremonti2007}, finding blueshifted absorption in 4/13 post-starburst galaxies. \citet{Maltby2019} observed blueshifted outflows in Mg II in a stacked sample of 40 post-starbursts at $1<z<1.4$, which may have been launched by either AGN activity or the recent starburst.

Statistical evidence for gas depletion during the post-starburst phase has been seen in studies of the gas and dust masses evolving with time. \citet{French2018} found a trend in the molecular gas to stellar mass ratios and post-starburst age in combining the molecular gas samples of \citet{French2015,Rowlands2015,Alatalo2016b}, with the molecular gas declining on $\sim 100-200$ Myr timescales. \citet{Li2019} observed a similar decline timescale in the dust masses for samples of post-starbursts with archival IR photometry. \citet{Davis2018} found post-starburst galaxies selected from the EAGLE simulation to follow a similar trend. These implied molecular gas depletion rates of $2-150$ M$_\odot$ yr$^{-1}$ are too rapid to be driven by the residual current star formation rates, and may be caused by AGN feedback.

The various roles of AGN feedback may work in combination with one another. Turbulence from low level AGN activity may be capable of suppressing star formation in the remaining molecular gas reservoirs left over after a starburst \citep{Alatalo2014, Smercina2018}, while outflows deplete the gas over $\sim 100-200$ Myr timescales \citep{French2018,Davis2018,Li2019}. High resolution spatially and kinematically resolved observations will shed light on these possible mechanisms.

\subsection{Environmental Effects}

While post-starburst galaxies are often observed in clusters, most post-starburst galaxies are in the field or in poor groups \citep{Zabludoff1996}, with slightly lower abundances in low-density regions and slightly higher abundances in clusters or with close neighbors relative to the population of star forming galaxies \citep{Hogg2006}. The relative importance of clusters in the post-starburst population may depend on redshift \citep{Yan2009a, delucia2009, Webb2020}. The post-starburst fraction rises sharply with environmental density, from $\lesssim 1$\% in the field to $\sim15$\% in dense clusters \citep{Balogh1999,Poggianti2009,Socolovsky2018,Paccagnella2019} (although, see \citealt{vonderLinden2010}).

Galaxies in clusters can be affected by both the cluster environment and by pre-processing from infalling galaxy groups \citep{Zabludoff1998}. Galaxies in dense cluster environments can undergo additional processes that result in the rapid end of star formation, producing a post-starburst spectral signature. These processes include ram pressure stripping, galaxy harassment, thermal evaporation, and starvation \citep[see review by][]{Boselli2006}. Post-starburst galaxies in clusters are found near the cluster centers, and with high clustercentric velocities, in a ``ring" of phase space consistent with a rapid end to star-formation in these galaxies upon first infall \citep{Muzzin2014, Paccagnella2017, Owers2019}. The rapid cessation of star formation, as well as results from spatially resolved studies of cluster post-starbursts are consistent with ram pressure stripping ending star formation in these galaxies \citep{Vulcani2020}. Post-starburst galaxies in simulations show different properties depending on their environment \citep{Wilkinson2018}, suggesting different mechanisms are at work for galaxies evolving through this phase in denser environments. Detailed comparisons of the ISM in post-starburst galaxies in different environments will shed further light on which mechanisms have acted to end star formation in these galaxies.

\subsection{Other Mechanisms}

For galaxies at higher redshift, high gas fractions can lead to a starburst via compaction \citep{Zolotov2015}. An episode of compaction could be triggered by a merger or period of rapid gas accretion. This additional mechanism for starbursts may contribute to the higher rate of post-starburst galaxies at higher redshift (see further discussion in \S\ref{sec:cosmictime}).

Cosmic ray feedback may also play a role in ending star formation in these galaxies. \citet{Owen2019} model the role of heating from high energy cosmic rays in high redshift ($7<z<9$) galaxies, finding cosmic ray heating capable of suppressing star formation and preventing future star formation by preventing gas inflows.

Many secular processes have been proposed to end star formation in galaxies without feedback or external events. Morphological quenching \citep{Martig2013,Gensior2020} can occur when a galaxy builds up its central bulge over time, which can ultimately act to stabilize the gas in a galaxy against collapse into stars. The timescale for morphological quenching is slow however, of order several Gyr, longer than the timescale for post-starburst galaxies to have ended their starbursts ($<200$ Myr, see \S\ref{sec:stellarpops}). Halo quenching \citep{Dekel2006} may also act to end star formation in galaxies, preventing further gas accretion after galaxy halos reach masses $>10^{12}$ M$_\odot$ by shock heating any infalling gas. While these mechanisms may not end star formation in the progenitors of post-starburst galaxies, these mechanisms may act to keep galaxies quiescent after star formation has ended.

\section{The future: evolution to quiescence or future star formation}
\label{sec:future}

The structural changes experienced by post-starburst galaxies sets them on a path to evolve to normal early type galaxies within a few Gyr. \citet{Yang2008} found that given time for the stellar populations to age and tidal features to fade, post-starburst galaxies will have morphological and kinematic properties consistent with early types in $\sim 3-7$ Gyr. The observed decline in gas and dust during this phase by \citet{French2018,Davis2018,Li2019} will result in early type levels of gas and dust after $\sim 1-2$ Gyr. 

Will these galaxies stay quiescent? Some galaxies will have grown their bulges to a degree where preventative feedback mechanisms are now effective. Longer duration mechanisms such as halo quenching \citep{Dekel2006} or morphological quenching \citep{Martig2013, Gensior2020} may act to prevent further star formation. Left-over gas stabilized against further star formation could form stars at a later time after being disturbed by a merger or interaction.

Other galaxies may experience further star formation if new gas is brought in via minor gas-rich mergers. Observations of misaligned, low metallicity gas in early type galaxies \citep{vandevoort2018, Davis2019} show that low level star formation can occur in early types when this happens. \citet{Dressler2013,Abramson2013} find the spatial density of post-starburst galaxies to be more closely related to quiescent galaxies, instead of starburst galaxies, and proposed a picture where many starbursts will return to normal star formation instead of going through a post-starburst quiescent phase, and many post-starburst galaxies have come from gas accretion onto previously quiescent galaxies. Some post-starburst galaxies may be experiencing a temporary halt in star formation, which will later resume. \citet{Pawlik2018,Pawlik2019} suggests this is the case for $\sim30-40$\% of post-starburst galaxies, selected using a PCA analysis similar to that of \citet{Wild2007}. \citet{Yamauchi2008} find companion galaxies around 8\% of post-starburst galaxies from the \citet{Goto2007} catalog, suggesting that some post-starburst galaxies may be observed after an initial burst of star formation but before the coalescence phase of a merger. In these cases, the system may experience another burst of star formation upon the final coalescence phase depending on the gas properties and geometry of the merger.

The fate of post-starburst galaxies will likely depend on their stellar masses, redshift, morphologies, and the type of starburst they experienced. Further work connecting stellar populations in progenitor sequences across cosmic time, and in understanding the onset of preventative feedback will answer lingering questions about the role of the post-starburst phase in galaxy evolution.

\section{Open questions and future facilities}

Recent work has used the post-starburst phase to address a number of puzzles in the field of galaxy evolution, as summarized in this review. Despite the rarity of these galaxies, the short duration of this phase and its increasing prevalence with redshift imply this is a common phase of galaxy evolution \citep{Wild2009, Snyder2011, Wild2016, Rowlands2018, Belli2019,Wild2020}. Yet there are a number of outstanding questions that remain:

\begin{enumerate}
    \item How do starbursts and their subsequent end proceed spatially within post-starburst galaxies? What sets the stellar population gradients observed in post-starburst galaxies?
    \item How do the gas reservoirs of galaxies change as they evolve to quiescence?
    \item What does the circumgalactic medium around post-starburst galaxies tell us about outflows and the possibility of future inflows?
    \item What is the role of AGN during the post-starburst phase? How often does AGN activity capable of disrupting or preventing star formation occur during this phase?
    \item How do the processes which rapidly end star formation in galaxies change with redshift, stellar mass, and environment?
    \item How do the mechanisms that end star formation rapidly compare to mechanisms that end star formation more gradually?
    \item How do galaxies begin preventative feedback after becoming quiescent?
\end{enumerate}

Future astronomical facilities will play a large role in addressing these questions. \textit{JWST} and large ground-based surveys like MOONS on the VLT can provide rest-frame optical observations of post-starburst and quiescent galaxies out to higher redshifts, and at lower stellar masses than previously accessible, as well as provide more accurate SFR and AGN indicators at lower redshifts. 30-m class ground-based telescopes will be able to constrain the central kinematics and resolved stellar populations of these galaxies to understand the connection between structural and stellar population properties in large numbers of galaxies. The next generation VLA would allow for molecular gas observations of higher redshift post-starburst galaxies in enough detail to constrain the detailed state of this gas and connect to various feedback mechanisms. 

\acknowledgements
KDF thanks Vivienne Wild, Francois Schweizer, Kate Rowlands, and Ann Zabludoff for their helpful feedback and comments on a draft of this review. KDF thanks the referee for their useful feedback, which has improved this review.

\bibliographystyle{aasjournal}
\bibliography{refs}

\begin{thebibliography}{}
\expandafter\ifx\csname natexlab\endcsname\relax\def\natexlab#1{#1}\fi
\providecommand{\url}[1]{\href{#1}{#1}}

\bibitem[{{Abramson} {et~al.}(2013){Abramson}, {Dressler}, {Gladders},
  {Oemler}, {Poggianti}, {Monson}, {Persson}, \& {Vulcani}}]{Abramson2013}
{Abramson}, L.~E., {Dressler}, A., {Gladders}, M.~D., {et~al.} 2013, \apj, 777,
  124

\bibitem[{Alatalo {et~al.}(2016{\natexlab{a}})Alatalo, Aladro, Nyland, Aalto,
  Bitsakis, Gallagher, \& Lanz}]{Alatalo2016a}
Alatalo, K., Aladro, R., Nyland, K., {et~al.} 2016{\natexlab{a}}, The
  Astrophysical Journal, 830, 137

\bibitem[{{Alatalo} {et~al.}(2011){Alatalo}, {Blitz}, {Young}, {Davis},
  {Bureau}, {Lopez}, {Cappellari}, {Scott}, {Shapiro}, {Crocker},
  {Mart{\'\i}n}, {Bois}, {Bournaud}, {Davies}, {de Zeeuw}, {Duc}, {Emsellem},
  {Falc{\'o}n-Barroso}, {Khochfar}, {Krajnovi{\'c}}, {Kuntschner}, {Lablanche},
  {McDermid}, {Morganti}, {Naab}, {Oosterloo}, {Sarzi}, {Serra}, \&
  {Weijmans}}]{Alatalo2011}
{Alatalo}, K., {Blitz}, L., {Young}, L.~M., {et~al.} 2011, \apj, 735, 88

\bibitem[{Alatalo {et~al.}(2014)Alatalo, Nyland, Graves, Griffin, Duc,
  Cappellari, McDermid, Davis, Crocker, Young, Chang, Scott, Cales, Bayet,
  Blitz, Bois, Bournaud, Bureau, Davies, de~Zeeuw, Emsellem, Khochfar,
  Krajnovi{\'{c}}, Kuntschner, Morganti, Naab, Oosterloo, Sarzi, Serra,
  Weijmans, \& Weijmans}]{Alatalo2014}
Alatalo, K., Nyland, K., Graves, G., {et~al.} 2014, The Astrophysical Journal,
  780, 11

\bibitem[{Alatalo {et~al.}(2015)Alatalo, Lacy, Lanz, Bitsakis, Appleton,
  Nyland, Cales, Chang, Davis, de~Zeeuw, Lonsdale, Mart{\'{i}}n, Meier, \&
  Ogle}]{Alatalo2015}
Alatalo, K., Lacy, M., Lanz, L., {et~al.} 2015, The Astrophysical Journal, 798,
  31

\bibitem[{Alatalo {et~al.}(2016{\natexlab{b}})Alatalo, Cales, Rich, Appleton,
  Kewley, Lacy, Lanz, Medling, \& Nyland}]{Alatalo2016}
Alatalo, K., Cales, S.~L., Rich, J.~A., {et~al.} 2016{\natexlab{b}}, The
  Astrophysical Journal Supplement Series, 224, 38

\bibitem[{Alatalo {et~al.}(2016{\natexlab{c}})Alatalo, Lisenfeld, Lanz,
  Appleton, Cales, Kewley, Lacy, Medling, Nyland, Rich, Urry, \&
  Urry}]{Alatalo2016b}
Alatalo, K., Lisenfeld, U., Lanz, L., {et~al.} 2016{\natexlab{c}}, The
  Astrophysical Journal, 827, 106

\bibitem[{Alatalo {et~al.}(2017)Alatalo, Bitsakis, Lanz, Lacy, Brown, French,
  Ciesla, Appleton, Beaton, Cales, Crossett, Falc{\'{o}}n-Barroso, Kelson,
  Kewley, Kriek, Medling, Mulchaey, Nyland, Rich, Urry, Bitsakis, Brown,
  Ciesla, Appleton, Beaton, Cales, Crossett, Falcon-Barroso, French, Kewley,
  Kelson, Kriek, Lanz, Medling, Mulchaey, Nyland, Rich, \& Urry}]{Alatalo2016c}
Alatalo, K., Bitsakis, T., Lanz, L., {et~al.} 2017, Astrophysical Journal, 843,
  9

\bibitem[{{Almaini} {et~al.}(2017){Almaini}, {Wild}, {Maltby}, {Hartley},
  {Simpson}, {Hatch}, {McLure}, {Dunlop}, \& {Rowlands}}]{Almaini2017}
{Almaini}, O., {Wild}, V., {Maltby}, D.~T., {et~al.} 2017, \mnras, 472, 1401

\bibitem[{Arcavi {et~al.}(2014)Arcavi, Gal-Yam, Sullivan, Pan, Cenko, Horesh,
  Ofek, {De Cia}, Yan, Yang, Howell, Tal, Kulkarni, Tendulkar, Tang, Xu,
  Sternberg, Cohen, Bloom, Nugent, Kasliwal, Perley, Quimby, Miller, Theissen,
  \& Laher}]{Arcavi2014}
Arcavi, I., Gal-Yam, A., Sullivan, M., {et~al.} 2014, The Astrophysical
  Journal, 793, 38

\bibitem[{{Arp}(1969)}]{Arp1969}
{Arp}, H. 1969, \aap, 3, 418

\bibitem[{{Baldwin} {et~al.}(1981){Baldwin}, {Phillips}, \&
  {Terlevich}}]{Baldwin1981}
{Baldwin}, J.~A., {Phillips}, M.~M., \& {Terlevich}, R. 1981, \pasp, 93, 5

\bibitem[{Balogh {et~al.}(1999)Balogh, Morris, Yee, Carlberg, \&
  Ellingson}]{Balogh1999}
Balogh, M.~L., Morris, S.~L., Yee, H. K.~C., Carlberg, R.~G., \& Ellingson, E.
  1999, The Astrophysical Journal, 527, 54

\bibitem[{{Barnes}(1988)}]{Barnes1988}
{Barnes}, J.~E. 1988, \apj, 331, 699

\bibitem[{{Barnes} \& {Hernquist}(1996)}]{Barnes1996}
{Barnes}, J.~E., \& {Hernquist}, L. 1996, \apj, 471, 115

\bibitem[{{Barnes} \& {Hernquist}(1991)}]{Barnes1991}
{Barnes}, J.~E., \& {Hernquist}, L.~E. 1991, \apjl, 370, L65

\bibitem[{{Baron} {et~al.}(2020){Baron}, {Netzer}, {Davies}, \& {Xavier
  Prochaska}}]{Baron2020}
{Baron}, D., {Netzer}, H., {Davies}, R.~I., \& {Xavier Prochaska}, J. 2020,
  \mnras, 494, 5396

\bibitem[{{Baron} {et~al.}(2021){Baron}, {Netzer}, {Lutz}, {Prochaska}, \&
  {Davies}}]{Baron2021}
{Baron}, D., {Netzer}, H., {Lutz}, D., {Prochaska}, J.~X., \& {Davies}, R.~I.
  2021, arXiv e-prints, arXiv:2105.08071

\bibitem[{Baron {et~al.}(2017)Baron, Netzer, Poznanski, Prochaska, \&
  Schreiber}]{Baron2017}
Baron, D., Netzer, H., Poznanski, D., Prochaska, J.~X., \& Schreiber, N. M.~F.
  2017, Monthly Notices of the Royal Astronomical Society, 470, 1687

\bibitem[{{Baron} {et~al.}(2017){Baron}, {Poznanski}, \& {}}]{Baron2017b}
{Baron}, D., {Poznanski}, D., \& {}. 2017, \mnras, 465, 4530

\bibitem[{{Baron} {et~al.}(2018){Baron}, {Netzer}, {Prochaska}, {Cai},
  {Cantalupo}, {Martin}, {Matuszewski}, {Moore}, {Morrissey}, \&
  {Neill}}]{Baron2018}
{Baron}, D., {Netzer}, H., {Prochaska}, J.~X., {et~al.} 2018, \mnras, 480, 3993

\bibitem[{{Becker} {et~al.}(1995){Becker}, {White}, \& {Helfand}}]{first}
{Becker}, R.~H., {White}, R.~L., \& {Helfand}, D.~J. 1995, \apj, 450, 559

\bibitem[{Bekki {et~al.}(2005)Bekki, Couch, Shioya, \& Vazdekis}]{Bekki2005}
Bekki, K., Couch, W.~J., Shioya, Y., \& Vazdekis, A. 2005, Monthly Notices of
  the Royal Astronomical Society, 359, 949

\bibitem[{{Belfiore} {et~al.}(2016){Belfiore}, {Maiolino}, {Maraston},
  {Emsellem}, {Bershady}, {Masters}, {Yan}, {Bizyaev}, {Boquien}, {Brownstein},
  {Bundy}, {Drory}, {Heckman}, {Law}, {Roman-Lopes}, {Pan}, {Stanghellini},
  {Thomas}, {Weijmans}, \& {Westfall}}]{Belfiore2016}
{Belfiore}, F., {Maiolino}, R., {Maraston}, C., {et~al.} 2016, \mnras, 461,
  3111

\bibitem[{Belli {et~al.}(2019)Belli, Newman, \& Ellis}]{Belli2019}
Belli, S., Newman, A.~B., \& Ellis, R.~S. 2019, The Astrophysical Journal, 874,
  17

\bibitem[{{Belli} {et~al.}(2021){Belli}, {Contursi}, {Genzel}, {Tacconi},
  {F{\"o}rster-Schreiber}, {Lutz}, {Combes}, {Neri}, {Garc{\'\i}a-Burillo},
  {Schuster}, {Herrera-Camus}, {Tadaki}, {Davies}, {Davies}, {Johnson}, {Lee},
  {Leja}, {Nelson}, {Price}, {Shangguan}, {Shimizu}, {Tacchella}, \&
  {{\"U}bler}}]{Belli2021}
{Belli}, S., {Contursi}, A., {Genzel}, R., {et~al.} 2021, \apjl, 909, L11

\bibitem[{{Blake} {et~al.}(2004){Blake}, {Pracy}, {Couch}, {Bekki}, {Lewis},
  {Glazebrook}, {Baldry}, {Baugh}, {Bland-Hawthorn}, {Bridges}, {Cannon},
  {Cole}, {Colless}, {Collins}, {Dalton}, {De Propris}, {Driver}, {Efstathiou},
  {Ellis}, {Frenk}, {Jackson}, {Lahav}, {Lumsden}, {Maddox}, {Madgwick},
  {Norberg}, {Peacock}, {Peterson}, {Sutherland}, \& {Taylor}}]{Blake2004}
{Blake}, C., {Pracy}, M.~B., {Couch}, W.~J., {et~al.} 2004, \mnras, 355, 713

\bibitem[{{Bolatto} {et~al.}(2013){Bolatto}, {Warren}, {Leroy}, {Walter},
  {Veilleux}, {Ostriker}, {Ott}, {Zwaan}, {Fisher}, {Weiss}, {Rosolowsky}, \&
  {Hodge}}]{Bolatto2013}
{Bolatto}, A.~D., {Warren}, S.~R., {Leroy}, A.~K., {et~al.} 2013, \nat, 499,
  450

\bibitem[{{Boselli} \& {Gavazzi}(2006)}]{Boselli2006}
{Boselli}, A., \& {Gavazzi}, G. 2006, \pasp, 118, 517

\bibitem[{Bravo-Alfaro {et~al.}(2001)Bravo-Alfaro, Cayatte, van Gorkom, \&
  Balkowski}]{Bravo?Alfaro2001}
Bravo-Alfaro, H., Cayatte, V., van Gorkom, J.~H., \& Balkowski, C. 2001,
  Astronomy and Astrophysics, 379, 347

\bibitem[{Brown {et~al.}(2009)Brown, Moustakas, Caldwell, Palamara, Cool, Dey,
  Hickox, Jannuzi, Murray, \& Zaritsky}]{Brown2009}
Brown, M. J.~I., Moustakas, J., Caldwell, N., {et~al.} 2009, The Astrophysical
  Journal, 703, 150

\bibitem[{{Bundy} {et~al.}(2015){Bundy}, {Bershady}, {Law}, {Yan}, {Drory},
  {MacDonald}, {Wake}, {Cherinka}, {S{\'a}nchez-Gallego}, {Weijmans}, {Thomas},
  {Tremonti}, {Masters}, {Coccato}, {Diamond-Stanic}, {Arag{\'o}n-Salamanca},
  {Avila-Reese}, {Badenes}, {Falc{\'o}n-Barroso}, {Belfiore}, {Bizyaev},
  {Blanc}, {Bland-Hawthorn}, {Blanton}, {Brownstein}, {Byler}, {Cappellari},
  {Conroy}, {Dutton}, {Emsellem}, {Etherington}, {Frinchaboy}, {Fu}, {Gunn},
  {Harding}, {Johnston}, {Kauffmann}, {Kinemuchi}, {Klaene}, {Knapen},
  {Leauthaud}, {Li}, {Lin}, {Maiolino}, {Malanushenko}, {Malanushenko}, {Mao},
  {Maraston}, {McDermid}, {Merrifield}, {Nichol}, {Oravetz}, {Pan}, {Parejko},
  {Sanchez}, {Schlegel}, {Simmons}, {Steele}, {Steinmetz}, {Thanjavur},
  {Thompson}, {Tinker}, {van den Bosch}, {Westfall}, {Wilkinson}, {Wright},
  {Xiao}, \& {Zhang}}]{manga}
{Bundy}, K., {Bershady}, M.~A., {Law}, D.~R., {et~al.} 2015, \apj, 798, 7

\bibitem[{Buyle {et~al.}(2008)Buyle, {De Rijcke}, \& Dejonghe}]{Buyle2008}
Buyle, P., {De Rijcke}, S., \& Dejonghe, H. 2008, The Astrophysical Journal,
  684, L17

\bibitem[{Buyle {et~al.}(2006)Buyle, Michielsen, {De Rijcke}, Pisano, Dejonghe,
  \& Freeman}]{Buyle2006}
Buyle, P., Michielsen, D., {De Rijcke}, S., {et~al.} 2006, The Astrophysical
  Journal, 649, 163

\bibitem[{Caldwell {et~al.}(1996)Caldwell, Rose, Franx, \&
  Leonardi}]{Caldwell1996}
Caldwell, N., Rose, J.~A., Franx, M., \& Leonardi, A.~J. 1996, The Astronomical
  Journal, 111, 78

\bibitem[{Cales \& Brotherton(2015)}]{Cales2015}
Cales, S.~L., \& Brotherton, M.~S. 2015, Monthly Notices of the Royal
  Astronomical Society, 449, 2374

\bibitem[{Cales {et~al.}(2013)Cales, Brotherton, Shang, Runnoe, DiPompeo,
  Bennert, Canalizo, Hiner, Stoll, Ganguly, \& Diamond-Stanic}]{Cales2013}
Cales, S.~L., Brotherton, M.~S., Shang, Z., {et~al.} 2013, The Astrophysical
  Journal, 762, 90

\bibitem[{Chang {et~al.}(2001)Chang, van Gorkom, Zabludoff, Zaritsky, \&
  Mihos}]{Chang2001}
Chang, T.-C., van Gorkom, J.~H., Zabludoff, A.~I., Zaritsky, D., \& Mihos,
  J.~C. 2001, The Astronomical Journal, 121, 1965

\bibitem[{{Chen} {et~al.}(2019){Chen}, {Shi}, {Wild}, {Tremonti}, {Rowlands},
  {Bizyaev}, {Yan}, {Lin}, \& {Riffel}}]{Chen2019}
{Chen}, Y.-M., {Shi}, Y., {Wild}, V., {et~al.} 2019, \mnras, 489, 5709

\bibitem[{{Cherinka} {et~al.}(2019){Cherinka}, {Andrews},
  {S{\'a}nchez-Gallego}, {Brownstein}, {Argudo-Fern{\'a}ndez}, {Blanton},
  {Bundy}, {Jones}, {Masters}, {Law}, {Rowlands}, {Weijmans}, {Westfall}, \&
  {Yan}}]{marvin}
{Cherinka}, B., {Andrews}, B.~H., {S{\'a}nchez-Gallego}, J., {et~al.} 2019,
  \aj, 158, 74

\bibitem[{{Cicone} {et~al.}(2014){Cicone}, {Maiolino}, {Sturm},
  {Graci{\'a}-Carpio}, {Feruglio}, {Neri}, {Aalto}, {Davies}, {Fiore},
  {Fischer}, {Garc{\'\i}a-Burillo}, {Gonz{\'a}lez-Alfonso}, {Hailey-Dunsheath},
  {Piconcelli}, \& {Veilleux}}]{Cicone2014}
{Cicone}, C., {Maiolino}, R., {Sturm}, E., {et~al.} 2014, \aap, 562, A21

\bibitem[{{Cid Fernandes} {et~al.}(2011){Cid Fernandes}, {Stasi{\'n}ska},
  {Mateus}, \& {Vale Asari}}]{CidFernandes2011}
{Cid Fernandes}, R., {Stasi{\'n}ska}, G., {Mateus}, A., \& {Vale Asari}, N.
  2011, \mnras, 413, 1687

\bibitem[{Ciesla {et~al.}(2016)Ciesla, Boselli, Elbaz, Boissier, Buat,
  Charmandaris, Schreiber, B{\'{e}}thermin, Baes, Boquien, {De Looze},
  Fern{\'{a}}ndez-Ontiveros, Pappalardo, Spinoglio, \& Viaene}]{Ciesla2016}
Ciesla, L., Boselli, A., Elbaz, D., {et~al.} 2016, Astronomy {\&} Astrophysics,
  585, 43

\bibitem[{{Citro} {et~al.}(2017){Citro}, {Pozzetti}, {Quai}, {Moresco},
  {Vallini}, \& {Cimatti}}]{Citro2017}
{Citro}, A., {Pozzetti}, L., {Quai}, S., {et~al.} 2017, \mnras, 469, 3108

\bibitem[{Coil {et~al.}(2011)Coil, Weiner, Holz, Cooper, Yan, \&
  Aird}]{Coil2011}
Coil, A.~L., Weiner, B.~J., Holz, D.~E., {et~al.} 2011, The Astrophysical
  Journal, 743, 46

\bibitem[{{Conroy}(2013)}]{Conroy2013}
{Conroy}, C. 2013, \araa, 51, 393

\bibitem[{{Conroy} \& {Gunn}(2010)}]{Conroy2010}
{Conroy}, C., \& {Gunn}, J.~E. 2010, \apj, 712, 833

\bibitem[{Couch \& Sharples(1987)}]{Couch1987}
Couch, W.~J., \& Sharples, R.~M. 1987, Monthly Notices of the Royal
  Astronomical Society, 229, 423

\bibitem[{{D'Agostino} {et~al.}(2019){D'Agostino}, {Kewley}, {Groves},
  {Medling}, {Di Teodoro}, {Dopita}, {Thomas}, {Sutherland}, \&
  {Garcia-Burillo}}]{DAgostino2019}
{D'Agostino}, J.~J., {Kewley}, L.~J., {Groves}, B.~A., {et~al.} 2019, \mnras,
  487, 4153

\bibitem[{{Davies} {et~al.}(2016){Davies}, {Groves}, {Kewley}, {Dopita},
  {Hampton}, {Shastri}, {Scharw{\"a}chter}, {Sutherland}, {Kharb}, {Bhatt},
  {Jin}, {Banfield}, {Zaw}, {James}, {Juneau}, \& {Srivastava}}]{Davies2016}
{Davies}, R.~L., {Groves}, B., {Kewley}, L.~J., {et~al.} 2016, \mnras, 462,
  1616

\bibitem[{{Davis} {et~al.}(2019){Davis}, {van de Voort}, {Rowlands},
  {McAlpine}, {Wild}, \& {Crain}}]{Davis2018}
{Davis}, T.~A., {van de Voort}, F., {Rowlands}, K., {et~al.} 2019, \mnras, 484,
  2447

\bibitem[{{Davis} \& {Young}(2019)}]{Davis2019}
{Davis}, T.~A., \& {Young}, L.~M. 2019, \mnras, 489, L108

\bibitem[{{De Lucia} {et~al.}(2009){De Lucia}, {Poggianti}, {Halliday},
  {Milvang-Jensen}, {Noll}, {Smail}, \& {Zaritsky}}]{delucia2009}
{De Lucia}, G., {Poggianti}, B.~M., {Halliday}, C., {et~al.} 2009, \mnras, 400,
  68

\bibitem[{{De Propris} \& Melnick(2014)}]{DePropris2014}
{De Propris}, R., \& Melnick, J. 2014, Monthly Notices of the Royal
  Astronomical Society, 439, 2837

\bibitem[{{Dekel} \& {Birnboim}(2006)}]{Dekel2006}
{Dekel}, A., \& {Birnboim}, Y. 2006, \mnras, 368, 2

\bibitem[{{D'Eugenio} {et~al.}(2020){D'Eugenio}, {Daddi}, {Gobat},
  {Strazzullo}, {Lustig}, {Delvecchio}, {Jin}, {Puglisi}, {Calabr{\'o}},
  {Mancini}, {Dickinson}, {Cimatti}, \& {Onodera}}]{DEugenio2020b}
{D'Eugenio}, C., {Daddi}, E., {Gobat}, R., {et~al.} 2020, \apjl, 892, L2

\bibitem[{D'Eugenio {et~al.}(2020)D'Eugenio, van~der Wel, Wu, Barone, van
  Houdt, Bezanson, Straatman, Pacifici, Muzzin, Gallazzi, Wild, Sobral, Bell,
  Zibetti, Mowla, \& Franx}]{DEugenio2020}
D'Eugenio, F., van~der Wel, A., Wu, P.-F., {et~al.} 2020, MNRAS, 000, 1

\bibitem[{Dressler \& Gunn(1983)}]{Dressler1983}
Dressler, A., \& Gunn, J.~E. 1983, The Astrophysical Journal, 270, 7

\bibitem[{{Dressler} {et~al.}(2013){Dressler}, {Oemler}, {Poggianti},
  {Gladders}, {Abramson}, \& {Vulcani}}]{Dressler2013}
{Dressler}, A., {Oemler}, Augustus, J., {Poggianti}, B.~M., {et~al.} 2013,
  \apj, 770, 62

\bibitem[{Dressler {et~al.}(1999)Dressler, Smail, Poggianti, Butcher, Couch,
  Ellis, \& {Oemler, Jr.}}]{Dressler1999}
Dressler, A., Smail, I., Poggianti, B.~M., {et~al.} 1999, The Astrophysical
  Journal Supplement Series, 122, 51

\bibitem[{{D'Souza} \& {Bell}(2018)}]{DSouza2018}
{D'Souza}, R., \& {Bell}, E.~F. 2018, Nature Astronomy, 2, 737

\bibitem[{Du {et~al.}(2010)Du, Luo, Prugniel, Liang, \& Zhao}]{Du2010a}
Du, W., Luo, A.~L., Prugniel, P., Liang, Y.~C., \& Zhao, Y.~H. 2010, Monthly
  Notices of the Royal Astronomical Society, 409, 567

\bibitem[{Dupraz {et~al.}(1990)Dupraz, Casoli, Combes, \& Kazes}]{Dupraz1990}
Dupraz, C., Casoli, F., Combes, F., \& Kazes, I. 1990, Astronomy and
  Astrophysics (ISSN 0004-6361), 228

\bibitem[{{Eldridge} \& {Stanway}(2020)}]{Eldridge2020}
{Eldridge}, J.~J., \& {Stanway}, E.~R. 2020, arXiv e-prints, arXiv:2005.11883

\bibitem[{{Emsellem} {et~al.}(2007){Emsellem}, {Cappellari}, {Krajnovi{\'c}},
  {van de Ven}, {Bacon}, {Bureau}, {Davies}, {de Zeeuw}, {Falc{\'o}n-Barroso},
  {Kuntschner}, {McDermid}, {Peletier}, \& {Sarzi}}]{Emsellem2007}
{Emsellem}, E., {Cappellari}, M., {Krajnovi{\'c}}, D., {et~al.} 2007, \mnras,
  379, 401

\bibitem[{{Emsellem} {et~al.}(2011){Emsellem}, {Cappellari}, {Krajnovi{\'c}},
  {Alatalo}, {Blitz}, {Bois}, {Bournaud}, {Bureau}, {Davies}, {Davis}, {de
  Zeeuw}, {Khochfar}, {Kuntschner}, {Lablanche}, {McDermid}, {Morganti},
  {Naab}, {Oosterloo}, {Sarzi}, {Scott}, {Serra}, {van de Ven}, {Weijmans}, \&
  {Young}}]{Emsellem2011}
---. 2011, \mnras, 414, 888

\bibitem[{{Faber} {et~al.}(1997){Faber}, {Tremaine}, {Ajhar}, {Byun},
  {Dressler}, {Gebhardt}, {Grillmair}, {Kormendy}, {Lauer}, \&
  {Richstone}}]{Faber1997}
{Faber}, S.~M., {Tremaine}, S., {Ajhar}, E.~A., {et~al.} 1997, \aj, 114, 1771

\bibitem[{{Feruglio} {et~al.}(2010){Feruglio}, {Maiolino}, {Piconcelli},
  {Menci}, {Aussel}, {Lamastra}, \& {Fiore}}]{Feruglio2010}
{Feruglio}, C., {Maiolino}, R., {Piconcelli}, E., {et~al.} 2010, \aap, 518,
  L155

\bibitem[{{Forrest} {et~al.}(2020){Forrest}, {Marsan}, {Annunziatella},
  {Wilson}, {Muzzin}, {Marchesini}, {Cooper}, {Chan}, {McConachie}, {Gomez},
  {Kado-Fong}, {Barbera}, {Lange-Vagle}, {Nantais}, {Nonino}, {Saracco},
  {Stefanon}, \& {van der Burg}}]{Forrest2020}
{Forrest}, B., {Marsan}, Z.~C., {Annunziatella}, M., {et~al.} 2020, \apj, 903,
  47

\bibitem[{{French} {et~al.}(2016){French}, {Arcavi}, \&
  {Zabludoff}}]{French2016}
{French}, K.~D., {Arcavi}, I., \& {Zabludoff}, A. 2016, \apjl, 818, L21

\bibitem[{{French} {et~al.}(2017){French}, {Arcavi}, \&
  {Zabludoff}}]{French2017}
---. 2017, \apj, 835, 176

\bibitem[{{French} {et~al.}(2020{\natexlab{a}}){French}, {Arcavi}, {Zabludoff},
  {Stone}, {Hiramatsu}, {van Velzen}, {McCully}, \& {Jiang}}]{French2020b}
{French}, K.~D., {Arcavi}, I., {Zabludoff}, A.~I., {et~al.} 2020{\natexlab{a}},
  \apj, 891, 93

\bibitem[{{French} {et~al.}(2020{\natexlab{b}}){French}, {Wevers}, {Law-Smith},
  {Graur}, \& {Zabludoff}}]{French2020}
{French}, K.~D., {Wevers}, T., {Law-Smith}, J., {Graur}, O., \& {Zabludoff},
  A.~I. 2020{\natexlab{b}}, \ssr, 216, 32

\bibitem[{French {et~al.}(2015)French, Yang, Zabludoff, Narayanan, Shirley,
  Walter, Smith, \& Tremonti}]{French2015}
French, K.~D., Yang, Y., Zabludoff, A., {et~al.} 2015, The Astrophysical
  Journal, 801, 1

\bibitem[{French {et~al.}(2018{\natexlab{a}})French, Yang, Zabludoff, \&
  Tremonti}]{French2018}
French, K.~D., Yang, Y., Zabludoff, A.~I., \& Tremonti, C.~A.
  2018{\natexlab{a}}, The Astrophysical Journal, 862, 2

\bibitem[{{French} \& {Zabludoff}(2018)}]{French2018c}
{French}, K.~D., \& {Zabludoff}, A.~I. 2018, \apj, 868, 99

\bibitem[{French {et~al.}(2018{\natexlab{b}})French, Zabludoff, Yoon, Shirley,
  Yang, Smercina, Smith, \& Narayanan}]{French2018b}
French, K.~D., Zabludoff, A.~I., Yoon, I., {et~al.} 2018{\natexlab{b}}, The
  Astrophysical Journal, 861, 123

\bibitem[{{Gao} \& {Solomon}(2004)}]{Gao2004}
{Gao}, Y., \& {Solomon}, P.~M. 2004, \apj, 606, 271

\bibitem[{{Geach} {et~al.}(2014){Geach}, {Hickox}, {Diamond-Stanic}, {Krips},
  {Rudnick}, {Tremonti}, {Sell}, {Coil}, \& {Moustakas}}]{Geach2014}
{Geach}, J.~E., {Hickox}, R.~C., {Diamond-Stanic}, A.~M., {et~al.} 2014, \nat,
  516, 68

\bibitem[{{Geach} {et~al.}(2018){Geach}, {Tremonti}, {Diamond-Stanic}, {Sell},
  {Kepley}, {Coil}, {Rudnick}, {Hickox}, {Moustakas}, \& {Yang}}]{Geach2018}
{Geach}, J.~E., {Tremonti}, C., {Diamond-Stanic}, A.~M., {et~al.} 2018, \apjl,
  864, L1

\bibitem[{{Gensior} {et~al.}(2020){Gensior}, {Kruijssen}, \&
  {Keller}}]{Gensior2020}
{Gensior}, J., {Kruijssen}, J.~M.~D., \& {Keller}, B.~W. 2020, \mnras, 495, 199

\bibitem[{{Georgakakis} {et~al.}(2008){Georgakakis}, {Nandra}, {Yan},
  {Willner}, {Lotz}, {Pierce}, {Cooper}, {Laird}, {Koo}, {Barmby}, {Newman},
  {Primack}, \& {Coil}}]{Georgakakis2008}
{Georgakakis}, A., {Nandra}, K., {Yan}, R., {et~al.} 2008, \mnras, 385, 2049

\bibitem[{{Goto}(2005)}]{Goto2005}
{Goto}, T. 2005, \mnras, 357, 937

\bibitem[{Goto(2007)}]{Goto2007}
Goto, T. 2007, Monthly Notices of the Royal Astronomical Society, 377, 1222

\bibitem[{Goto {et~al.}(2003)Goto, Nichol, Okamura, Sekiguchi, Miller,
  Bernardi, Hopkins, Tremonti, Connolly, Castander, Brinkmann, Fukugita,
  Harvanek, Ivezi{\'{c}}, Kleinman, Krzesinski, Long, Loveday, Neilsen, Newman,
  Nitta, Snedden, \& SubbaRao}]{Goto2003}
Goto, T., Nichol, R.~C., Okamura, S., {et~al.} 2003, Publications of the
  Astronomical Society of Japan, 55, 771

\bibitem[{{Graur} {et~al.}(2018){Graur}, {French}, {Zahid}, {Guillochon},
  {Mandel}, {Auchettl}, \& {Zabludoff}}]{Graur2018}
{Graur}, O., {French}, K.~D., {Zahid}, H.~J., {et~al.} 2018, \apj, 853, 39

\bibitem[{{Greene} {et~al.}(2020){Greene}, {Setton}, {Bezanson}, {Suess},
  {Kriek}, {Spilker}, {Goulding}, \& {Feldmann}}]{Greene2020}
{Greene}, J.~E., {Setton}, D., {Bezanson}, R., {et~al.} 2020, \apjl, 899, L9

\bibitem[{{Hayward} {et~al.}(2014){Hayward}, {Lanz}, {Ashby}, {Fazio},
  {Hernquist}, {Mart{\'\i}nez-Galarza}, {Noeske}, {Smith}, {Wuyts}, \&
  {Zezas}}]{Hayward2014}
{Hayward}, C.~C., {Lanz}, L., {Ashby}, M. L.~N., {et~al.} 2014, \mnras, 445,
  1598

\bibitem[{{Heckman} {et~al.}(2017){Heckman}, {Borthakur}, {Wild},
  {Schiminovich}, \& {Bordoloi}}]{Heckman2017}
{Heckman}, T., {Borthakur}, S., {Wild}, V., {Schiminovich}, D., \& {Bordoloi},
  R. 2017, \apj, 846, 151

\bibitem[{Hogg {et~al.}(2006)Hogg, Masjedi, Berlind, Blanton, Quintero, \&
  Brinkmann}]{Hogg2006}
Hogg, D.~W., Masjedi, M., Berlind, A.~A., {et~al.} 2006, The Astrophysical
  Journal, 650, 763

\bibitem[{Hopkins(2012)}]{Hopkins2012b}
Hopkins, P.~F. 2012, Monthly Notices of the Royal Astronomical Society:
  Letters, 420, L8

\bibitem[{{Hopkins} {et~al.}(2009){Hopkins}, {Cox}, {Dutta}, {Hernquist},
  {Kormendy}, \& {Lauer}}]{Hopkins2009}
{Hopkins}, P.~F., {Cox}, T.~J., {Dutta}, S.~N., {et~al.} 2009, \apjs, 181, 135

\bibitem[{{Hopkins} {et~al.}(2013){Hopkins}, {Cox}, {Hernquist}, {Narayanan},
  {Hayward}, \& {Murray}}]{Hopkins2013}
{Hopkins}, P.~F., {Cox}, T.~J., {Hernquist}, L., {et~al.} 2013, \mnras, 430,
  1901

\bibitem[{{Hopkins} {et~al.}(2006){Hopkins}, {Hernquist}, {Cox}, {Di Matteo},
  {Robertson}, \& {Springel}}]{Hopkins2006}
{Hopkins}, P.~F., {Hernquist}, L., {Cox}, T.~J., {et~al.} 2006, \apjs, 163, 1

\bibitem[{{Hopkins} {et~al.}(2012){Hopkins}, {Quataert}, \&
  {Murray}}]{Hopkins2012}
{Hopkins}, P.~F., {Quataert}, E., \& {Murray}, N. 2012, \mnras, 421, 3522

\bibitem[{Hunt {et~al.}(2018)Hunt, Bezanson, Greene, Spilker, Suess, Kriek,
  Narayanan, Feldmann, van~der Wel, \& Pattarakijwanich}]{Hunt2018}
Hunt, Q., Bezanson, R., Greene, J.~E., {et~al.} 2018, The Astrophysical
  Journal, 860, L18

\bibitem[{{Johnson} {et~al.}(2020){Johnson}, {Leja}, {Conroy}, \&
  {Speagle}}]{Johnson2017}
{Johnson}, B.~D., {Leja}, J., {Conroy}, C., \& {Speagle}, J.~S. 2020, arXiv
  e-prints, arXiv:2012.01426

\bibitem[{{Juneau} {et~al.}(2009){Juneau}, {Narayanan}, {Moustakas}, {Shirley},
  {Bussmann}, {Kennicutt}, \& {Vanden Bout}}]{Juneau2009}
{Juneau}, S., {Narayanan}, D.~T., {Moustakas}, J., {et~al.} 2009, \apj, 707,
  1217

\bibitem[{{Kauffmann} {et~al.}(2003){Kauffmann}, {Heckman}, {Tremonti},
  {Brinchmann}, {Charlot}, {White}, {Ridgway}, {Brinkmann}, {Fukugita}, {Hall},
  {Ivezi{\'c}}, {Richards}, \& {Schneider}}]{Kauffmann2003}
{Kauffmann}, G., {Heckman}, T.~M., {Tremonti}, C., {et~al.} 2003, \mnras, 346,
  1055

\bibitem[{Kaviraj {et~al.}(2007)Kaviraj, Kirkby, Silk, \& Sarzi}]{Kaviraj2007a}
Kaviraj, S., Kirkby, L.~A., Silk, J., \& Sarzi, M. 2007, Monthly Notices of the
  Royal Astronomical Society, 382, 960

\bibitem[{{Keel} {et~al.}(2012){Keel}, {Chojnowski}, {Bennert}, {Schawinski},
  {Lintott}, {Lynn}, {Pancoast}, {Harris}, {Nierenberg}, {Sonnenfeld}, \&
  {Proctor}}]{Keel2012}
{Keel}, W.~C., {Chojnowski}, S.~D., {Bennert}, V.~N., {et~al.} 2012, \mnras,
  420, 878

\bibitem[{{Kennicutt}(1998)}]{Kennicutt1998}
{Kennicutt}, Robert~C., J. 1998, \apj, 498, 541

\bibitem[{Kohno {et~al.}(2002)Kohno, Tosaki, Matsushita, Vila-Vila{\'{o}},
  Shibatsuka, \& Kawabe}]{Kohno2002}
Kohno, K., Tosaki, T., Matsushita, S., {et~al.} 2002, Publications of the
  Astronomical Society of Japan, 54, 541

\bibitem[{Kriek {et~al.}(2010)Kriek, Labb{\'{e}}, Conroy, Whitaker, van Dokkum,
  Brammer, Franx, Illingworth, Marchesini, Muzzin, Quadri, \&
  Rudnick}]{Kriek2010}
Kriek, M., Labb{\'{e}}, I., Conroy, C., {et~al.} 2010, The Astrophysical
  Journal Letters, 722, L64

\bibitem[{{Kruijssen} {et~al.}(2019){Kruijssen}, {Schruba}, {Chevance},
  {Longmore}, {Hygate}, {Haydon}, {McLeod}, {Dalcanton}, {Tacconi}, \& {van
  Dishoeck}}]{Kruijssen2019}
{Kruijssen}, J.~M.~D., {Schruba}, A., {Chevance}, M., {et~al.} 2019, \nat, 569,
  519

\bibitem[{{Larson} \& {Tinsley}(1978)}]{Larson1978}
{Larson}, R.~B., \& {Tinsley}, B.~M. 1978, \apj, 219, 46

\bibitem[{{Lauer} {et~al.}(1995){Lauer}, {Ajhar}, {Byun}, {Dressler}, {Faber},
  {Grillmair}, {Kormendy}, {Richstone}, \& {Tremaine}}]{Lauer1995}
{Lauer}, T.~R., {Ajhar}, E.~A., {Byun}, Y.~I., {et~al.} 1995, \aj, 110, 2622

\bibitem[{{Law} {et~al.}(2020){Law}, {Belfiore}, {Ji}, {Bershady},
  {Cappellari}, {Westfall}, {Yan}, {Bizyaev}, {Brownstein}, {Drory}, \&
  {Andrews}}]{Law2020}
{Law}, D.~R., {Belfiore}, F., {Ji}, X., {et~al.} 2020, arXiv e-prints,
  arXiv:2011.06012

\bibitem[{{Law-Smith} {et~al.}(2017){Law-Smith}, {Ramirez-Ruiz}, {Ellison}, \&
  {Foley}}]{Law-Smith2017}
{Law-Smith}, J., {Ramirez-Ruiz}, E., {Ellison}, S.~L., \& {Foley}, R.~J. 2017,
  \apj, 850, 22

\bibitem[{{Leitherer} {et~al.}(1999){Leitherer}, {Schaerer}, {Goldader},
  {Delgado}, {Robert}, {Kune}, {de Mello}, {Devost}, \& {Heckman}}]{sb99}
{Leitherer}, C., {Schaerer}, D., {Goldader}, J.~D., {et~al.} 1999, \apjs, 123,
  3

\bibitem[{Leonardi \& Rose(1996)}]{Leonardi1996}
Leonardi, A.~J., \& Rose, J.~A. 1996, The Astronomical Journal, 111, 182

\bibitem[{Li {et~al.}(2019)Li, French, Zabludoff, \& Ho}]{Li2019}
Li, Z., French, K.~D., Zabludoff, A.~I., \& Ho, L.~C. 2019, The Astrophysical
  Journal, 879, 131

\bibitem[{{Lintott} {et~al.}(2009){Lintott}, {Schawinski}, {Keel}, {van Arkel},
  {Bennert}, {Edmondson}, {Thomas}, {Smith}, {Herbert}, {Jarvis}, {Virani},
  {Andreescu}, {Bamford}, {Land}, {Murray}, {Nichol}, {Raddick}, {Slosar},
  {Szalay}, \& {Vandenberg}}]{Lintott2009}
{Lintott}, C.~J., {Schawinski}, K., {Keel}, W., {et~al.} 2009, \mnras, 399, 129

\bibitem[{Liu \& Green(1996)}]{Liu1996}
Liu, C.~T., \& Green, R.~F. 1996, The Astrophysical Journal Letters, 458, L63

\bibitem[{{Lotz} {et~al.}(2008){Lotz}, {Jonsson}, {Cox}, \&
  {Primack}}]{Lotz2008}
{Lotz}, J.~M., {Jonsson}, P., {Cox}, T.~J., \& {Primack}, J.~R. 2008, \mnras,
  391, 1137

\bibitem[{{Maltby} {et~al.}(2018){Maltby}, {Almaini}, {Wild}, {Hatch},
  {Hartley}, {Simpson}, {Rowlands}, \& {Socolovsky}}]{Maltby2018}
{Maltby}, D.~T., {Almaini}, O., {Wild}, V., {et~al.} 2018, \mnras, 480, 381

\bibitem[{Maltby {et~al.}(2016)Maltby, Almaini, Wild, Hatch, Hartley, Simpson,
  McLure, Dunlop, Rowlands, \& Cirasuolo}]{Maltby2016}
Maltby, D.~T., Almaini, O., Wild, V., {et~al.} 2016, Monthly Notices of the
  Royal Astronomical Society: Letters, 459, L114

\bibitem[{Maltby {et~al.}(2019)Maltby, Almaini, McLure, Wild, Dunlop, Rowlands,
  Hartley, Hatch, Socolovsky, Wilkinson, Amorin, Bradshaw, Carnall, Castellano,
  Cimatti, Cresci, Cullen, {De Barros}, Fontanot, Garilli, Koekemoer, McLeod,
  Pentericci, \& Talia}]{Maltby2019}
Maltby, D.~T., Almaini, O., McLure, R.~J., {et~al.} 2019, Monthly Notices of
  the Royal Astronomical Society, 489, 1139

\bibitem[{{Martig} {et~al.}(2013){Martig}, {Crocker}, {Bournaud}, {Emsellem},
  {Gabor}, {Alatalo}, {Blitz}, {Bois}, {Bureau}, {Cappellari}, {Davies},
  {Davis}, {Dekel}, {de Zeeuw}, {Duc}, {Falc{\'o}n-Barroso}, {Khochfar},
  {Krajnovi{\'c}}, {Kuntschner}, {Morganti}, {McDermid}, {Naab}, {Oosterloo},
  {Sarzi}, {Scott}, {Serra}, {Griffin}, {Teyssier}, {Weijmans}, \&
  {Young}}]{Martig2013}
{Martig}, M., {Crocker}, A.~F., {Bournaud}, F., {et~al.} 2013, \mnras, 432,
  1914

\bibitem[{{Matharu} {et~al.}(2020){Matharu}, {Muzzin}, {Brammer}, {van der
  Burg}, {Auger}, {Hewett}, {Chan}, {Demarco}, {van Dokkum}, {Marchesini},
  {Nelson}, {Noble}, \& {Wilson}}]{Matharu2019}
{Matharu}, J., {Muzzin}, A., {Brammer}, G.~B., {et~al.} 2020, \mnras, 493, 6011

\bibitem[{{Matsushita} {et~al.}(2010){Matsushita}, {Kawabe}, {Kohno}, {Tosaki},
  \& {Vila-Vilar{\'o}}}]{Matsushita2010}
{Matsushita}, S., {Kawabe}, R., {Kohno}, K., {Tosaki}, T., \&
  {Vila-Vilar{\'o}}, B. 2010, \pasj, 62, 409

\bibitem[{{McQuinn} {et~al.}(2010){McQuinn}, {Skillman}, {Cannon}, {Dalcanton},
  {Dolphin}, {Hidalgo-Rodr{\'\i}guez}, {Holtzman}, {Stark}, {Weisz}, \&
  {Williams}}]{McQuinn2010}
{McQuinn}, K. B.~W., {Skillman}, E.~D., {Cannon}, J.~M., {et~al.} 2010, \apj,
  721, 297

\bibitem[{Melnick \& {De Propris}(2013)}]{Melnick2013}
Melnick, J., \& {De Propris}, R. 2013, Monthly Notices of the Royal
  Astronomical Society, 431, 2034

\bibitem[{Mendel {et~al.}(2013)Mendel, Simard, Ellison, \& Patton}]{Mendel2013}
Mendel, J.~T., Simard, L., Ellison, S.~L., \& Patton, D.~R. 2013, Monthly
  Notices of the Royal Astronomical Society, 429, 2212

\bibitem[{{Meusinger} {et~al.}(2017){Meusinger}, {Br{\"u}necke}, {Schalldach},
  \& {in der Au}}]{Meusinger2016}
{Meusinger}, H., {Br{\"u}necke}, J., {Schalldach}, P., \& {in der Au}, A. 2017,
  \aap, 597, A134

\bibitem[{{Mihos} \& {Hernquist}(1994)}]{Mihos1994}
{Mihos}, J.~C., \& {Hernquist}, L. 1994, \apjl, 431, L9

\bibitem[{Miller \& Owen(2001)}]{Miller2001}
Miller, N.~A., \& Owen, F.~N. 2001, The Astrophysical Journal, 554, L25

\bibitem[{{Mori{\'c}} {et~al.}(2010){Mori{\'c}}, {Smol{\v{c}}i{\'c}},
  {Kimball}, {Riechers}, {Ivezi{\'c}}, \& {Scoville}}]{Moric2010}
{Mori{\'c}}, I., {Smol{\v{c}}i{\'c}}, V., {Kimball}, A., {et~al.} 2010, \apj,
  724, 779

\bibitem[{{Morishita} {et~al.}(2021){Morishita}, {D'Amato}, {Abramson},
  {Abdurro'uf}, {Stiavelli}, \& {Lucas}}]{Morishita2021}
{Morishita}, T., {D'Amato}, Q., {Abramson}, L.~E., {et~al.} 2021, \apj, 908,
  163

\bibitem[{{Muzzin} {et~al.}(2014){Muzzin}, {van der Burg}, {McGee}, {Balogh},
  {Franx}, {Hoekstra}, {Hudson}, {Noble}, {Taranu}, {Webb}, {Wilson}, \&
  {Yee}}]{Muzzin2014}
{Muzzin}, A., {van der Burg}, R.~F.~J., {McGee}, S.~L., {et~al.} 2014, \apj,
  796, 65

\bibitem[{{Naab} \& {Burkert}(2003)}]{Naab2003}
{Naab}, T., \& {Burkert}, A. 2003, \apj, 597, 893

\bibitem[{{Naab} {et~al.}(2014){Naab}, {Oser}, {Emsellem}, {Cappellari},
  {Krajnovi{\'c}}, {McDermid}, {Alatalo}, {Bayet}, {Blitz}, {Bois}, {Bournaud},
  {Bureau}, {Crocker}, {Davies}, {Davis}, {de Zeeuw}, {Duc}, {Hirschmann},
  {Johansson}, {Khochfar}, {Kuntschner}, {Morganti}, {Oosterloo}, {Sarzi},
  {Scott}, {Serra}, {van de Ven}, {Weijmans}, \& {Young}}]{Naab2014}
{Naab}, T., {Oser}, L., {Emsellem}, E., {et~al.} 2014, \mnras, 444, 3357

\bibitem[{Nielsen {et~al.}(2012)Nielsen, Ridgway, {De Propris}, \&
  Goto}]{Nielsen2012}
Nielsen, D.~M., Ridgway, S.~E., {De Propris}, R., \& Goto, T. 2012, The
  Astrophysical Journal, 761, L16

\bibitem[{Norton {et~al.}(2001)Norton, Gebhardt, Zabludoff, \&
  Zaritsky}]{Norton2001}
Norton, S.~A., Gebhardt, K., Zabludoff, A.~I., \& Zaritsky, D. 2001, The
  Astrophysical Journal, 557, 150

\bibitem[{{Owen} {et~al.}(2019){Owen}, {Wu}, {Jin}, {Surajbali}, \&
  {Kataoka}}]{Owen2019}
{Owen}, E.~R., {Wu}, K., {Jin}, X., {Surajbali}, P., \& {Kataoka}, N. 2019,
  \aap, 626, A85

\bibitem[{{Owers} {et~al.}(2019){Owers}, {Hudson}, {Oman}, {Bland-Hawthorn},
  {Brough}, {Bryant}, {Cortese}, {Couch}, {Croom}, {van de Sande}, {Federrath},
  {Groves}, {Hopkins}, {Lawrence}, {Lorente}, {McDermid}, {Medling},
  {Richards}, {Scott}, {Taranu}, {Welker}, \& {Yi}}]{Owers2019}
{Owers}, M.~S., {Hudson}, M.~J., {Oman}, K.~A., {et~al.} 2019, \apj, 873, 52

\bibitem[{{Paccagnella} {et~al.}(2019){Paccagnella}, {Vulcani}, {Poggianti},
  {Moretti}, {Fritz}, {Gullieuszik}, \& {Fasano}}]{Paccagnella2019}
{Paccagnella}, A., {Vulcani}, B., {Poggianti}, B.~M., {et~al.} 2019, \mnras,
  482, 881

\bibitem[{{Paccagnella} {et~al.}(2017){Paccagnella}, {Vulcani}, {Poggianti},
  {Fritz}, {Fasano}, {Moretti}, {Jaff{\'e}}, {Biviano}, {Gullieuszik},
  {Bettoni}, {Cava}, {Couch}, \& {D'Onofrio}}]{Paccagnella2017}
---. 2017, \apj, 838, 148

\bibitem[{{Pattarakijwanich} {et~al.}(2016){Pattarakijwanich}, {Strauss}, {Ho},
  \& {Ross}}]{Pattarakijwanich2016}
{Pattarakijwanich}, P., {Strauss}, M.~A., {Ho}, S., \& {Ross}, N.~P. 2016,
  \apj, 833, 19

\bibitem[{{Pawlik} {et~al.}(2019){Pawlik}, {McAlpine}, {Trayford}, {Wild},
  {Bower}, {Crain}, {Schaller}, \& {Schaye}}]{Pawlik2019}
{Pawlik}, M.~M., {McAlpine}, S., {Trayford}, J.~W., {et~al.} 2019, Nature
  Astronomy, 3, 440

\bibitem[{Pawlik {et~al.}(2015)Pawlik, Wild, Walcher, Johansson, Villforth,
  Rowlands, Mendez-Abreu, \& Hewlett}]{Pawlik2015}
Pawlik, M.~M., Wild, V., Walcher, C.~J., {et~al.} 2015, Monthly Notices of the
  Royal Astronomical Society, 456, 3032

\bibitem[{{Pawlik} {et~al.}(2018){Pawlik}, {Taj Aldeen}, {Wild},
  {Mendez-Abreu}, {Lah{\'e}n}, {Johansson}, {Jimenez}, {Lucas}, {Zheng},
  {Walcher}, \& {Rowlands}}]{Pawlik2018}
{Pawlik}, M.~M., {Taj Aldeen}, L., {Wild}, V., {et~al.} 2018, \mnras, 477, 1708

\bibitem[{{Poggianti} \& {Wu}(2000)}]{Poggianti2000}
{Poggianti}, B.~M., \& {Wu}, H. 2000, \apj, 529, 157

\bibitem[{{Poggianti} {et~al.}(2009){Poggianti}, {Arag{\'o}n-Salamanca},
  {Zaritsky}, {De Lucia}, {Milvang-Jensen}, {Desai}, {Jablonka}, {Halliday},
  {Rudnick}, {Varela}, {Bamford}, {Best}, {Clowe}, {Noll}, {Saglia},
  {Pell{\'o}}, {Simard}, {von der Linden}, \& {White}}]{Poggianti2009}
{Poggianti}, B.~M., {Arag{\'o}n-Salamanca}, A., {Zaritsky}, D., {et~al.} 2009,
  \apj, 693, 112

\bibitem[{{Pontzen} {et~al.}(2017){Pontzen}, {Tremmel}, {Roth}, {Peiris},
  {Saintonge}, {Volonteri}, {Quinn}, \& {Governato}}]{Pontzen2017}
{Pontzen}, A., {Tremmel}, M., {Roth}, N., {et~al.} 2017, \mnras, 465, 547

\bibitem[{Pracy {et~al.}(2009)Pracy, Kuntschner, Couch, Blake, Bekki, \&
  Briggs}]{Pracy2009}
Pracy, M.~B., Kuntschner, H., Couch, W.~J., {et~al.} 2009, Monthly Notices of
  the Royal Astronomical Society, 396, 1349

\bibitem[{Pracy {et~al.}(2012)Pracy, Owers, Couch, Kuntschner, Bekki, Briggs,
  Lah, \& Zwaan}]{Pracy2012}
Pracy, M.~B., Owers, M.~S., Couch, W.~J., {et~al.} 2012, Monthly Notices of the
  Royal Astronomical Society, 420, 2232

\bibitem[{Pracy {et~al.}(2014)Pracy, Owers, Zwaan, Couch, Kuntschner, Croom, \&
  Sadler}]{Pracy2014a}
Pracy, M.~B., Owers, M.~S., Zwaan, M., {et~al.} 2014, Monthly Notices of the
  Royal Astronomical Society, 443, 388

\bibitem[{Pracy {et~al.}(2013)Pracy, Croom, Sadler, Couch, Kuntschner, Bekki,
  Owers, Zwaan, Turner, \& Bergmann}]{Pracy2013}
Pracy, M.~B., Croom, S., Sadler, E., {et~al.} 2013, Monthly Notices of the
  Royal Astronomical Society, 432, 3131

\bibitem[{{Prieto} {et~al.}(2016){Prieto}, {Kr{\"u}hler}, {Anderson},
  {Galbany}, {Kochanek}, {Aquino}, {Brown}, {Dong}, {F{\"o}rster}, {Holoien},
  {Kuncarayakti}, {Maureira}, {Rosales-Ortega}, {S{\'a}nchez}, {Shappee}, \&
  {Stanek}}]{Prieto2016}
{Prieto}, J.~L., {Kr{\"u}hler}, T., {Anderson}, J.~P., {et~al.} 2016, \apjl,
  830, L32

\bibitem[{Quintero {et~al.}(2004)Quintero, Hogg, Blanton, Schlegel, Eisenstein,
  Gunn, Brinkmann, Fukugita, Glazebrook, \& Goto}]{Quintero2004}
Quintero, A.~D., Hogg, D.~W., Blanton, M.~R., {et~al.} 2004, The Astrophysical
  Journal, 602, 190

\bibitem[{{Rich} {et~al.}(2015){Rich}, {Kewley}, \& {Dopita}}]{Rich2015}
{Rich}, J.~A., {Kewley}, L.~J., \& {Dopita}, M.~A. 2015, \apjs, 221, 28

\bibitem[{Roseboom {et~al.}(2009)Roseboom, Oliver, \& Farrah}]{Roseboom2009}
Roseboom, I.~G., Oliver, S., \& Farrah, D. 2009, The Astrophysical Journal,
  699, L1

\bibitem[{{Roth} {et~al.}(2020){Roth}, {van Velzen}, {Cenko}, \&
  {Mushotzky}}]{Roth2020}
{Roth}, N., {van Velzen}, S., {Cenko}, S.~B., \& {Mushotzky}, R.~F. 2020, arXiv
  e-prints, arXiv:2008.11231

\bibitem[{Rowlands {et~al.}(2015)Rowlands, Wild, Nesvadba, Sibthorpe, Mortier,
  Lehnert, \& da~Cunha}]{Rowlands2015}
Rowlands, K., Wild, V., Nesvadba, N., {et~al.} 2015, Monthly Notices of the
  Royal Astronomical Society, 448, 258

\bibitem[{{Rowlands} {et~al.}(2018{\natexlab{a}}){Rowlands}, {Wild}, {Bourne},
  {Bremer}, {Brough}, {Driver}, {Hopkins}, {Owers}, {Phillipps}, {Pimbblet},
  {Sansom}, {Wang}, {Alpaslan}, {Bland-Hawthorn}, {Colless}, {Holwerda}, \&
  {Taylor}}]{Rowlands2017}
{Rowlands}, K., {Wild}, V., {Bourne}, N., {et~al.} 2018{\natexlab{a}}, \mnras,
  473, 1168

\bibitem[{{Rowlands} {et~al.}(2018{\natexlab{b}}){Rowlands}, {Heckman}, {Wild},
  {Zakamska}, {Rodriguez-Gomez}, {Barrera-Ballesteros}, {Lotz}, {Thilker},
  {Andrews}, {Boquien}, {Brinkmann}, {Brownstein}, {Hwang}, \&
  {Smethurst}}]{Rowlands2018}
{Rowlands}, K., {Heckman}, T., {Wild}, V., {et~al.} 2018{\natexlab{b}}, \mnras,
  480, 2544

\bibitem[{{Sazonova} {et~al.}(2021){Sazonova}, {Alatalo}, {Rowlands},
  {Deustua}, {French}, {Heckman}, {Lanz}, {Lisenfeld}, {Luo}, {Medling},
  {Nyland}, {Otter}, {Petric}, {Snyder}, \& {Urry}}]{Sazonova2021}
{Sazonova}, E., {Alatalo}, K., {Rowlands}, K., {et~al.} 2021, arXiv e-prints,
  arXiv:2105.09956

\bibitem[{{Schaerer} {et~al.}(1999){Schaerer}, {Contini}, \&
  {Kunth}}]{Schaerer1999}
{Schaerer}, D., {Contini}, T., \& {Kunth}, D. 1999, \aap, 341, 399

\bibitem[{{Schweizer}(1978)}]{Schweizer1978}
{Schweizer}, F. 1978, in Structure and Properties of Nearby Galaxies, ed. E.~M.
  {Berkhuijsen} \& R.~{Wielebinski}, Vol.~77, 279

\bibitem[{{Schweizer}(1982)}]{Schweizer1982}
{Schweizer}, F. 1982, \apj, 252, 455

\bibitem[{Schweizer {et~al.}(2013)Schweizer, Seitzer, Kelson, Villanueva, \&
  Walth}]{Schweizer2013}
Schweizer, F., Seitzer, P., Kelson, D.~D., Villanueva, E.~V., \& Walth, G.~L.
  2013, The Astrophysical Journal, 773, 148

\bibitem[{{Sell} {et~al.}(2014){Sell}, {Tremonti}, {Hickox}, {Diamond-Stanic},
  {Moustakas}, {Coil}, {Williams}, {Rudnick}, {Robaina}, {Geach}, {Heinz}, \&
  {Wilcots}}]{Sell2014}
{Sell}, P.~H., {Tremonti}, C.~A., {Hickox}, R.~C., {et~al.} 2014, \mnras, 441,
  3417

\bibitem[{{Setton} {et~al.}(2020){Setton}, {Bezanson}, {Suess}, {Hunt},
  {Greene}, {Kriek}, {Spilker}, {Feldmann}, \& {Narayanan}}]{Setton2020}
{Setton}, D.~J., {Bezanson}, R., {Suess}, K.~A., {et~al.} 2020, \apj, 905, 79

\bibitem[{{Shectman} {et~al.}(1996){Shectman}, {Landy}, {Oemler}, {Tucker},
  {Lin}, {Kirshner}, \& {Schechter}}]{Shectman1996}
{Shectman}, S.~A., {Landy}, S.~D., {Oemler}, A., {et~al.} 1996, \apj, 470, 172

\bibitem[{{Smail} {et~al.}(1999){Smail}, {Morrison}, {Gray}, {Owen}, {Ivison},
  {Kneib}, \& {Ellis}}]{Smail1999}
{Smail}, I., {Morrison}, G., {Gray}, M.~E., {et~al.} 1999, \apj, 525, 609

\bibitem[{Smercina {et~al.}(2018)Smercina, Smith, Dale, French, Croxall,
  Zhukovska, Togi, Bell, Crocker, Draine, Jarrett, Tremonti, Yang, \&
  Zabludoff}]{Smercina2018}
Smercina, A., Smith, J. D.~T., Dale, D.~A., {et~al.} 2018, The Astrophysical
  Journal, 855, 51

\bibitem[{Snyder {et~al.}(2011)Snyder, Cox, Hayward, Hernquist, \&
  Jonsson}]{Snyder2011}
Snyder, G.~F., Cox, T.~J., Hayward, C.~C., Hernquist, L., \& Jonsson, P. 2011,
  The Astrophysical Journal, 741, 77

\bibitem[{{Socolovsky} {et~al.}(2018){Socolovsky}, {Almaini}, {Hatch}, {Wild},
  {Maltby}, {Hartley}, \& {Simpson}}]{Socolovsky2018}
{Socolovsky}, M., {Almaini}, O., {Hatch}, N.~A., {et~al.} 2018, \mnras, 476,
  1242

\bibitem[{{Spinrad}(1973)}]{Spinrad1973}
{Spinrad}, H. 1973, \apj, 182, 381

\bibitem[{{Stern} {et~al.}(2012){Stern}, {Assef}, {Benford}, {Blain}, {Cutri},
  {Dey}, {Eisenhardt}, {Griffith}, {Jarrett}, {Lake}, {Masci}, {Petty},
  {Stanford}, {Tsai}, {Wright}, {Yan}, {Harrison}, \& {Madsen}}]{Stern2012}
{Stern}, D., {Assef}, R.~J., {Benford}, D.~J., {et~al.} 2012, \apj, 753, 30

\bibitem[{{Stinson} {et~al.}(2007){Stinson}, {Dalcanton}, {Quinn}, {Kaufmann},
  \& {Wadsley}}]{Stinson2007}
{Stinson}, G.~S., {Dalcanton}, J.~J., {Quinn}, T., {Kaufmann}, T., \&
  {Wadsley}, J. 2007, \apj, 667, 170

\bibitem[{{Stone} {et~al.}(2018){Stone}, {Generozov}, {Vasiliev}, \&
  {Metzger}}]{Stone2018}
{Stone}, N.~C., {Generozov}, A., {Vasiliev}, E., \& {Metzger}, B.~D. 2018,
  \mnras, 480, 5060

\bibitem[{{Strauss} {et~al.}(2002){Strauss}, {Weinberg}, {Lupton}, {Narayanan},
  {Annis}, {Bernardi}, {Blanton}, {Burles}, {Connolly}, {Dalcanton}, {Doi},
  {Eisenstein}, {Frieman}, {Fukugita}, {Gunn}, {Ivezi{\'c}}, {Kent}, {Kim},
  {Knapp}, {Kron}, {Munn}, {Newberg}, {Nichol}, {Okamura}, {Quinn}, {Richmond},
  {Schlegel}, {Shimasaku}, {SubbaRao}, {Szalay}, {Vanden Berk}, {Vogeley},
  {Yanny}, {Yasuda}, {York}, \& {Zehavi}}]{Strauss2002}
{Strauss}, M.~A., {Weinberg}, D.~H., {Lupton}, R.~H., {et~al.} 2002, \aj, 124,
  1810

\bibitem[{Suess {et~al.}(2017)Suess, Bezanson, Spilker, Kriek, Greene,
  Feldmann, Hunt, \& Narayanan}]{Suess2017}
Suess, K.~A., Bezanson, R., Spilker, J.~S., {et~al.} 2017, The Astrophysical
  Journal Letters, 846, L14

\bibitem[{{Suess} {et~al.}(2020){Suess}, {Kriek}, {Price}, \&
  {Barro}}]{Suess2020}
{Suess}, K.~A., {Kriek}, M., {Price}, S.~H., \& {Barro}, G. 2020, \apjl, 899,
  L26

\bibitem[{{Suess} {et~al.}(2021){Suess}, {Kriek}, {Price}, \&
  {Barro}}]{Suess2021}
---. 2021, arXiv e-prints, arXiv:2101.05820

\bibitem[{Swinbank {et~al.}(2012)Swinbank, Balogh, Bower, Zabludoff, Lucey,
  McGee, Miller, \& Nichol}]{Swinbank2012}
Swinbank, A.~M., Balogh, M.~L., Bower, R.~G., {et~al.} 2012, Monthly Notices of
  the Royal Astronomical Society, 420, 672

\bibitem[{{Tinsley}(1968)}]{Tinsley1968}
{Tinsley}, B.~M. 1968, \apj, 151, 547

\bibitem[{{Tinsley} \& {Larson}(1979)}]{Tinsley1979}
{Tinsley}, B.~M., \& {Larson}, R.~B. 1979, \mnras, 186, 503

\bibitem[{{Toft} {et~al.}(2014){Toft}, {Smol{\v{c}}i{\'c}}, {Magnelli},
  {Karim}, {Zirm}, {Michalowski}, {Capak}, {Sheth}, {Schawinski}, {Krogager},
  {Wuyts}, {Sanders}, {Man}, {Lutz}, {Staguhn}, {Berta}, {Mccracken}, {Krpan},
  \& {Riechers}}]{Toft2014}
{Toft}, S., {Smol{\v{c}}i{\'c}}, V., {Magnelli}, B., {et~al.} 2014, \apj, 782,
  68

\bibitem[{{Toomre} \& {Toomre}(1972)}]{Toomre1972}
{Toomre}, A., \& {Toomre}, J. 1972, \apj, 178, 623

\bibitem[{Tremonti {et~al.}(2007)Tremonti, Moustakas, \&
  Diamond-Stanic}]{Tremonti2007}
Tremonti, C.~A., Moustakas, J., \& Diamond-Stanic, A.~M. 2007, The
  Astrophysical Journal Letters, 663, L77

\bibitem[{Tripp {et~al.}(2011)Tripp, Meiring, Prochaska, Willmer, Howk, Werk,
  Jenkins, Bowen, Lehner, Sembach, Thom, \& Tumlinson}]{Tripp2011}
Tripp, T.~M., Meiring, J.~D., Prochaska, J.~X., {et~al.} 2011, Science (New
  York, N.Y.), 334, 952

\bibitem[{{van de Voort} {et~al.}(2018){van de Voort}, {Davis}, {Matsushita},
  {Rowlands}, {Shabala}, {Allison}, {Ting}, {Sansom}, \& {van der
  Werf}}]{vandevoort2018}
{van de Voort}, F., {Davis}, T.~A., {Matsushita}, S., {et~al.} 2018, \mnras,
  476, 122

\bibitem[{{van der Wel} {et~al.}(2014){van der Wel}, {Franx}, {van Dokkum},
  {Skelton}, {Momcheva}, {Whitaker}, {Brammer}, {Bell}, {Rix}, {Wuyts},
  {Ferguson}, {Holden}, {Barro}, {Koekemoer}, {Chang}, {McGrath},
  {H{\"a}ussler}, {Dekel}, {Behroozi}, {Fumagalli}, {Leja}, {Lundgren},
  {Maseda}, {Nelson}, {Wake}, {Patel}, {Labb{\'e}}, {Faber}, {Grogin}, \&
  {Kocevski}}]{vanderwel2014}
{van der Wel}, A., {Franx}, M., {van Dokkum}, P.~G., {et~al.} 2014, \apj, 788,
  28

\bibitem[{{Veilleux} {et~al.}(2005){Veilleux}, {Cecil}, \&
  {Bland-Hawthorn}}]{Veilleux2005}
{Veilleux}, S., {Cecil}, G., \& {Bland-Hawthorn}, J. 2005, \araa, 43, 769

\bibitem[{{von der Linden} {et~al.}(2010){von der Linden}, {Wild}, {Kauffmann},
  {White}, \& {Weinmann}}]{vonderLinden2010}
{von der Linden}, A., {Wild}, V., {Kauffmann}, G., {White}, S. D.~M., \&
  {Weinmann}, S. 2010, \mnras, 404, 1231

\bibitem[{{Vulcani} {et~al.}(2020){Vulcani}, {Fritz}, {Poggianti}, {Bettoni},
  {Franchetto}, {Moretti}, {Gullieuszik}, {Jaff{\'e}}, {Biviano}, {Radovich},
  \& {Mingozzi}}]{Vulcani2020}
{Vulcani}, B., {Fritz}, J., {Poggianti}, B.~M., {et~al.} 2020, \apj, 892, 146

\bibitem[{{Watkins} {et~al.}(2018){Watkins}, {Mihos}, {Bershady}, \&
  {Harding}}]{Watkins2018}
{Watkins}, A.~E., {Mihos}, J.~C., {Bershady}, M., \& {Harding}, P. 2018, \apjl,
  858, L16

\bibitem[{{Webb} {et~al.}(2020){Webb}, {Balogh}, {Leja}, {van der Burg},
  {Rudnick}, {Muzzin}, {Boak}, {Cerulo}, {Gilbank}, {Lidman}, {Old},
  {Pintos-Castro}, {McGee}, {Shipley}, {Biviano}, {Chan}, {Cooper}, {De Lucia},
  {Demarco}, {Forrest}, {Jablonka}, {Kukstas}, {McCarthy}, {McNab}, {Nantais},
  {Noble}, {Poggianti}, {Reeves}, {Vulcani}, {Wilson}, {Yee}, \&
  {Zaritsky}}]{Webb2020}
{Webb}, K., {Balogh}, M.~L., {Leja}, J., {et~al.} 2020, \mnras, 498, 5317

\bibitem[{Whitaker {et~al.}(2012)Whitaker, Kriek, van Dokkum, Bezanson,
  Brammer, Franx, \& Labb{\'{e}}}]{Whitaker2012}
Whitaker, K.~E., Kriek, M., van Dokkum, P.~G., {et~al.} 2012, The Astrophysical
  Journal, 745, 179

\bibitem[{Wild {et~al.}(2016)Wild, Almaini, Dunlop, Simpson, Rowlands, Bowler,
  Maltby, \& McLure}]{Wild2016}
Wild, V., Almaini, O., Dunlop, J., {et~al.} 2016, Monthly Notices of the Royal
  Astronomical Society, 463, 832

\bibitem[{Wild {et~al.}(2010)Wild, Heckman, \& Charlot}]{Wild2010}
Wild, V., Heckman, T., \& Charlot, S. 2010, Monthly Notices of the Royal
  Astronomical Society, 405, 933

\bibitem[{Wild {et~al.}(2007)Wild, Kauffmann, Heckman, Charlot, Lemson,
  Brinchmann, Reichard, \& Pasquali}]{Wild2007}
Wild, V., Kauffmann, G., Heckman, T., {et~al.} 2007, Monthly Notices of the
  Royal Astronomical Society, 381, 543

\bibitem[{Wild {et~al.}(2009)Wild, Walcher, Johansson, Tresse, Charlot, Pollo,
  {Le F{\`{e}}vre}, \& de~Ravel}]{Wild2009}
Wild, V., Walcher, C.~J., Johansson, P.~H., {et~al.} 2009, Monthly Notices of
  the Royal Astronomical Society, 395, 144

\bibitem[{Wild {et~al.}(2014)Wild, Almaini, Cirasuolo, Dunlop, McLure, Bowler,
  Ferreira, Bradshaw, Chuter, \& Hartley}]{Wild2014}
Wild, V., Almaini, O., Cirasuolo, M., {et~al.} 2014, Monthly Notices of the
  Royal Astronomical Society, 440, 1880

\bibitem[{{Wild} {et~al.}(2020){Wild}, {Taj Aldeen}, {Carnall}, {Maltby},
  {Almaini}, {Werle}, {Wilkinson}, {Rowlands}, {Bolzonella}, {Castellano},
  {Gargiulo}, {McLure}, {Pentericci}, \& {Pozzetti}}]{Wild2020}
{Wild}, V., {Taj Aldeen}, L., {Carnall}, A., {et~al.} 2020, \mnras, 494, 529

\bibitem[{{Wilkinson} {et~al.}(2021){Wilkinson}, {Almaini}, {Wild}, {Maltby},
  {Hartley}, {Simpson}, \& {Rowlands}}]{Wilkinson2021}
{Wilkinson}, A., {Almaini}, O., {Wild}, V., {et~al.} 2021, \mnras,
  arXiv:2104.07676

\bibitem[{{Wilkinson} {et~al.}(2018){Wilkinson}, {Pimbblet}, {Stott}, {Few}, \&
  {Gibson}}]{Wilkinson2018}
{Wilkinson}, C.~L., {Pimbblet}, K.~A., {Stott}, J.~P., {Few}, C.~G., \&
  {Gibson}, B.~K. 2018, \mnras, 479, 758

\bibitem[{{Williams} {et~al.}(2021){Williams}, {Spilker}, {Whitaker},
  {Dav{\'e}}, {Woodrum}, {Brammer}, {Bezanson}, {Narayanan}, \&
  {Weiner}}]{Williams2021}
{Williams}, C.~C., {Spilker}, J.~S., {Whitaker}, K.~E., {et~al.} 2021, \apj,
  908, 54

\bibitem[{Wong {et~al.}(2012)Wong, Schawinski, Kaviraj, Masters, Nichol,
  Lintott, Keel, Darg, Bamford, Andreescu, Murray, Raddick, Szalay, Thomas, \&
  VandenBerg}]{Wong2012}
Wong, O.~I., Schawinski, K., Kaviraj, S., {et~al.} 2012, Monthly Notices of the
  Royal Astronomical Society, 420, 1684

\bibitem[{{Worthey} \& {Ottaviani}(1997)}]{Worthey1997}
{Worthey}, G., \& {Ottaviani}, D.~L. 1997, \apjs, 111, 377

\bibitem[{{Wu} {et~al.}(2018){Wu}, {van der Wel}, {Bezanson}, {Gallazzi},
  {Pacifici}, {Straatman}, {Bari{\v{s}}i{\'c}}, {Bell}, {Chauke}, {van Houdt},
  {Franx}, {Muzzin}, {Sobral}, \& {Wild}}]{Wu2018}
{Wu}, P.-F., {van der Wel}, A., {Bezanson}, R., {et~al.} 2018, \apj, 868, 37

\bibitem[{{Yamauchi} {et~al.}(2008){Yamauchi}, {Yagi}, \&
  {Goto}}]{Yamauchi2008}
{Yamauchi}, C., {Yagi}, M., \& {Goto}, T. 2008, \mnras, 390, 383

\bibitem[{{Yan} \& {Blanton}(2012)}]{Yan2012}
{Yan}, R., \& {Blanton}, M.~R. 2012, \apj, 747, 61

\bibitem[{Yan {et~al.}(2006)Yan, Newman, Faber, Konidaris, Koo, \&
  Davis}]{Yan2006}
Yan, R., Newman, J.~A., Faber, S.~M., {et~al.} 2006, The Astrophysical Journal,
  648, 281

\bibitem[{Yan {et~al.}(2009)Yan, Newman, Faber, Coil, Cooper, Davis, Weiner,
  Gerke, \& Koo}]{Yan2009a}
---. 2009, Monthly Notices of the Royal Astronomical Society, 398, 735

\bibitem[{Yang {et~al.}(2006)Yang, Tremonti, Zabludoff, \& Zaritsky}]{Yang2006}
Yang, Y., Tremonti, C.~A., Zabludoff, A.~I., \& Zaritsky, D. 2006, The
  Astrophysical Journal, 646, L33

\bibitem[{Yang {et~al.}(2004)Yang, Zabludoff, Zaritsky, Lauer, \&
  Mihos}]{Yang2004}
Yang, Y., Zabludoff, A.~I., Zaritsky, D., Lauer, T.~R., \& Mihos, J.~C. 2004,
  The Astrophysical Journal, 607, 258

\bibitem[{Yang {et~al.}(2008)Yang, Zabludoff, Zaritsky, \& Mihos}]{Yang2008}
Yang, Y., Zabludoff, A.~I., Zaritsky, D., \& Mihos, J.~C. 2008, The
  Astrophysical Journal, 688, 945

\bibitem[{{Yano} {et~al.}(2016){Yano}, {Kriek}, {van der Wel}, \&
  {Whitaker}}]{Yano2016}
{Yano}, M., {Kriek}, M., {van der Wel}, A., \& {Whitaker}, K.~E. 2016, \apjl,
  817, L21

\bibitem[{Yesuf {et~al.}(2014)Yesuf, Faber, Trump, Koo, Fang, Liu, Wild, \&
  Hayward}]{Yesuf2014}
Yesuf, H.~M., Faber, S.~M., Trump, J.~R., {et~al.} 2014, The Astrophysical
  Journal, 792, 84

\bibitem[{{Yesuf} \& {Ho}(2020)}]{Yesuf2020}
{Yesuf}, H.~M., \& {Ho}, L.~C. 2020, \apj, 900, 107

\bibitem[{{York} {et~al.}(2000){York}, {Adelman}, {Anderson}, {Anderson},
  {Annis}, {Bahcall}, {Bakken}, {Barkhouser}, {Bastian}, {Berman}, {Boroski},
  {Bracker}, {Briegel}, {Briggs}, {Brinkmann}, {Brunner}, {Burles}, {Carey},
  {Carr}, {Castander}, {Chen}, {Colestock}, {Connolly}, {Crocker}, {Csabai},
  {Czarapata}, {Davis}, {Doi}, {Dombeck}, {Eisenstein}, {Ellman}, {Elms},
  {Evans}, {Fan}, {Federwitz}, {Fiscelli}, {Friedman}, {Frieman}, {Fukugita},
  {Gillespie}, {Gunn}, {Gurbani}, {de Haas}, {Haldeman}, {Harris}, {Hayes},
  {Heckman}, {Hennessy}, {Hindsley}, {Holm}, {Holmgren}, {Huang}, {Hull},
  {Husby}, {Ichikawa}, {Ichikawa}, {Ivezi{\'c}}, {Kent}, {Kim}, {Kinney},
  {Klaene}, {Kleinman}, {Kleinman}, {Knapp}, {Korienek}, {Kron}, {Kunszt},
  {Lamb}, {Lee}, {Leger}, {Limmongkol}, {Lindenmeyer}, {Long}, {Loomis},
  {Loveday}, {Lucinio}, {Lupton}, {MacKinnon}, {Mannery}, {Mantsch}, {Margon},
  {McGehee}, {McKay}, {Meiksin}, {Merelli}, {Monet}, {Munn}, {Narayanan},
  {Nash}, {Neilsen}, {Neswold}, {Newberg}, {Nichol}, {Nicinski}, {Nonino},
  {Okada}, {Okamura}, {Ostriker}, {Owen}, {Pauls}, {Peoples}, {Peterson},
  {Petravick}, {Pier}, {Pope}, {Pordes}, {Prosapio}, {Rechenmacher}, {Quinn},
  {Richards}, {Richmond}, {Rivetta}, {Rockosi}, {Ruthmansdorfer}, {Sandford},
  {Schlegel}, {Schneider}, {Sekiguchi}, {Sergey}, {Shimasaku}, {Siegmund},
  {Smee}, {Smith}, {Snedden}, {Stone}, {Stoughton}, {Strauss}, {Stubbs},
  {SubbaRao}, {Szalay}, {Szapudi}, {Szokoly}, {Thakar}, {Tremonti}, {Tucker},
  {Uomoto}, {Vanden Berk}, {Vogeley}, {Waddell}, {Wang}, {Watanabe},
  {Weinberg}, {Yanny}, {Yasuda}, \& {SDSS Collaboration}}]{York2000}
{York}, D.~G., {Adelman}, J., {Anderson}, John~E., J., {et~al.} 2000, \aj, 120,
  1579

\bibitem[{{Young} {et~al.}(2011){Young}, {Bureau}, {Davis}, {Combes},
  {McDermid}, {Alatalo}, {Blitz}, {Bois}, {Bournaud}, {Cappellari}, {Davies},
  {de Zeeuw}, {Emsellem}, {Khochfar}, {Krajnovi{\'c}}, {Kuntschner},
  {Lablanche}, {Morganti}, {Naab}, {Oosterloo}, {Sarzi}, {Scott}, {Serra}, \&
  {Weijmans}}]{Young2011}
{Young}, L.~M., {Bureau}, M., {Davis}, T.~A., {et~al.} 2011, \mnras, 414, 940

\bibitem[{{Zabludoff} \& {Mulchaey}(1998)}]{Zabludoff1998}
{Zabludoff}, A.~I., \& {Mulchaey}, J.~S. 1998, \apj, 496, 39

\bibitem[{Zabludoff {et~al.}(1996)Zabludoff, Zaritsky, Lin, Tucker, Hashimoto,
  Shectman, Oemler, \& Kirshner}]{Zabludoff1996}
Zabludoff, A.~I., Zaritsky, D., Lin, H., {et~al.} 1996, The Astrophysical
  Journal, 466, 104

\bibitem[{{Zahedy} {et~al.}(2020){Zahedy}, {Chen}, {Boettcher}, {Rauch},
  {Decker French}, \& {Zabludoff}}]{Zahedy2020}
{Zahedy}, F.~S., {Chen}, H.-W., {Boettcher}, E., {et~al.} 2020, \apjl, 904, L10

\bibitem[{{Zheng} {et~al.}(2020){Zheng}, {Wild}, {Lah{\'e}n}, {Johansson},
  {Law}, {Weaver}, \& {Jimenez}}]{Zheng2020}
{Zheng}, Y., {Wild}, V., {Lah{\'e}n}, N., {et~al.} 2020, \mnras, 498, 1259

\bibitem[{{Zibetti} {et~al.}(2013){Zibetti}, {Gallazzi}, {Charlot}, {Pierini},
  \& {Pasquali}}]{Zibetti2013}
{Zibetti}, S., {Gallazzi}, A., {Charlot}, S., {Pierini}, D., \& {Pasquali}, A.
  2013, \mnras, 428, 1479

\bibitem[{{Zolotov} {et~al.}(2015){Zolotov}, {Dekel}, {Mandelker}, {Tweed},
  {Inoue}, {DeGraf}, {Ceverino}, {Primack}, {Barro}, \& {Faber}}]{Zolotov2015}
{Zolotov}, A., {Dekel}, A., {Mandelker}, N., {et~al.} 2015, \mnras, 450, 2327

\bibitem[{Zwaan {et~al.}(2013)Zwaan, Kuntschner, Pracy, \& Couch}]{Zwaan2013}
Zwaan, M.~A., Kuntschner, H., Pracy, M.~B., \& Couch, W.~J. 2013, Monthly
  Notices of the Royal Astronomical Society, 432, 8

\end{thebibliography}

\end{document}